\documentclass[useAMS, usenatbib]{mnras}

\usepackage{newtxtext,newtxmath}

\usepackage[T1]{fontenc}
\usepackage{ae,aecompl}

\usepackage{graphicx}	
\usepackage{amsmath}	
\usepackage{amssymb}	

\usepackage{epsfig}
\usepackage{subfigure}
\usepackage{multirow}

\def \aj{Astronom.~J.}

\def \aap{A\&A }
\def \apjl{ApJ}
\def \apjs{ApJS}
\def \apj{ApJ}

\def \jqsrt{J.~Quant.~Spectrosc.~Radiat.~Transfer}

\def \mnras{MNRAS}
\def \prl{Phys.~Rev.~Lett.,}
\def \prd{Phys.~Rev.~D,}

\def \nat{Nature\ }

\newcommand{\msun}{M_{\odot}}

\newcommand{\fg}{f_\mathrm{\rm gas}}

\newcommand{\MA}{\mathcal{M}_{\rm A}}

\DeclareMathOperator{\sech}{sech}

\newcommand{\appropto}{\mathrel{\vcenter{
  \offinterlineskip\halign{\hfil$##$\cr
    \propto\cr\noalign{\kern2pt}\sim\cr\noalign{\kern-2pt}}}}}

\def\alt{\raise0.3ex\hbox{$\;<$\kern-0.75em\raise-1.1ex\hbox{$\sim\;$}}}
\def\agt{\raise0.3ex\hbox{$\;>$\kern-0.75em\raise-1.1ex\hbox{$\sim\;$}}}

\newcommand{\kms}{${\rm km~s^\mathrm{-1}}$}

\newcommand{\bw}{\begin{widetext}}
\newcommand{\ew}{\end{widetext}}

\newcommand{\lsim}{\,\rlap{\raise 0.35ex\hbox{$<$}}{\lower 0.7ex\hbox{$\sim$}}\,}
\newcommand{\gsim}{\,\rlap{\raise 0.35ex\hbox{$>$}}{\lower 0.7ex\hbox{$\sim$}}\,}

\defcitealias{Crocker2018}{CKTC18}
\defcitealias{Crocker2020a}{Paper I}

\newcommand{\aref}[1]{\hyperref[#1]{Appendix~\ref{#1}}}

\title[Cosmic Ray Feedback II]{
Cosmic rays across the star-forming galaxy sequence. II: Stability limits and the onset of cosmic ray-driven outflows
}

\author[R. M.~Crocker et al.]{
Roland M.~Crocker,$^{1}$\thanks{E-mail: rcrocker@fastmail.fm (RMC)}
Mark R. Krumholz,$^{1}$
and
Todd A. Thompson,$^{2}$
\\
$^{1}$Research School of Astronomy and Astrophysics, Australian National University, Canberra 2611, A.C.T., Australia\\
$^{2}$Department of Astronomy and Center for Cosmology \& Astro-Particle Physics, The Ohio State University, Columbus, Ohio 43210, U.S.A
}

\date{Accepted XXX. Received YYY; in original form ZZZ}

\pubyear{2020}

\begin{document}
\label{firstpage}
\pagerange{\pageref{firstpage}--\pageref{lastpage}}
\maketitle

\begin{abstract}
Cosmic rays (CRs) are a plausible mechanism for launching winds of cool material from the discs of star-forming galaxies. However, there is no consensus on what types of galaxies likely host CR-driven winds, or what role these winds might play in regulating galaxies' star formation rates. Using a detailed treatment of the
transport and losses of hadronic CRs developed in the previous paper in this series, here we develop a semi-analytic model that allows us to assess the viability of using CRs to launch cool winds from galactic discs. In particular, we determine the critical CR fluxes -- and corresponding star formation rate surface densities -- above which hydrostatic equilibrium within a given galaxy is precluded because CRs drive the gas off in a wind or otherwise render it unstable. We show that, 
for star-forming galaxies with lower gas surface densities typical of the Galaxy and local dwarfs, the locus of this CR stability curve patrols the high side of the observed distribution of galaxies in the Kennicutt-Schmidt parameter space of star formation rate versus gas surface density. However, hadronic losses render CRs unable to drive winds in galaxies with higher surface densities. Our results show that quiescent, low surface density galaxies like the Milky Way are poised on the cusp of instability, such that small changes to ISM parameters can lead to the launching of CR-driven outflows, and we suggest that, as a result, CR feedback sets an ultimate limit to the star formation efficiency of most modern galaxies.
\end{abstract}




\section{Introduction}

This paper is the third in a series \citep[][hereafter Paper I]{Krumholz2019,Crocker2020a} exploring 
the physics of relativistic
cosmic ray (CR) transport, energy loss, and radiation in the interstellar media of star-forming galaxies, and, more importantly,
the dynamical impact of CRs in such environments.
In particular, our intention in this series is to investigate, in broad brush strokes, the potential importance of CRs as an agent of {\it feedback} 
in star-forming galaxies:
What role, if any, do CRs have -- as a function of environmental parameters -- in establishing the remarkably low efficiency with which galaxies convert into stars the
gas flowing out of the cosmic web and into their own interstellar media?

As discussed in \citetalias{Crocker2020a} and previously literature, CRs are a plausible
agent of star formation feedback for a number of reasons:
While this non-thermal particle population receives only a sub-dominant fraction, $\sim 10\%$,  of the total kinetic energy liberated in supernova explosions, unlike the thermal gas (that receives most of the supernova energy), CRs lose energy to radiation very slowly.
This means that, from their injection sites close to the midplanes of star-forming galaxies, CRs tend to disperse well out into these galaxies' interstellar media\footnote{
In fact, in many cases, including for the Milky Way, they may escape the galactic disc completely.}.
Within the Milky Way disc, 
the CR energy density is near equipartition with the magnetic field and turbulent gas motions, implying CRs  contribute significantly to establishing the vertical hydrostatic equilibrium of the gas \citep[e.g.,][]{Boulares1990} and maintain, therefore, the conditions under which
sustained, quiescent star-formation can proceed.

Moreover, given their soft  effective equation of state\footnote{That follows from the fact that 
the energetically dominant cosmic ray population is relativistic, i.e.,
adiabatic index
$\gamma_c \to 4/3$.},
CRs come to increasingly dominate the total energy density of
a co-mingled astrophysical fluid of thermal and non-thermal particles that is suffering adiabatic losses under expansion in an outflow.
Thus CRs can help sustain galactic winds by providing a distributed heating source via their non-adiabatic energy losses which, in this situation,
are mostly mediated by the streaming instability
\citep[e.g.,][]{Everett2008,Zweibel2017,Ruszkowski2017}.
Despite, however, the early recognition of their potential importance 
in driving winds
\citep{Ipavich1975,Breitschwerdt1991,Zirakashvili1996,Ptuskin1997},
the possibility that CRs might generically be an important source of feedback in galaxy formation 
has only recently begun to receive much 
sustained attention, in either
phenomenological \citep[e.g.,][]{Zirakashvili2006, Everett2008,Samui2010,Crocker2011,Crocker2012, Lacki2011, Hanasz2013, Yoast-Hull2016}, or numerical models
\citep[e.g.,][]{Jubelgas2008,Wadepuhl2011,Uhlig2012,Booth2013,Salem2014,Salem2016,Pakmor2016,Simpson2016,Recchia2016,Recchia2017,Ruszkowski2017,Pfrommer2017,Chan2019,Buck2019}.
Even so, there remains significant disagreement in the literature about where and when CRs might be important: some authors conclude they are capable of driving galactic winds only off the most rapidly star-forming galaxies 
\citep[e.g.,][]{Socrates2008}, while others find they drive winds only in dwarfs, \citep[e.g.,][]{Jubelgas2008, Uhlig2012}, and yet others that they do not drive winds by themselves at all, but can reheat and energise winds launched by other processes \citep[e.g.][]{Ruszkowski2017}. 

Thus, a first-principles effort to understand where and when CRs might be important, taking into account all the available observational constraints, seems warranted, and this is the primary goal of this and our previous paper. 
Having explored the theory and observational consequences of CR transport in the largely neutral gas phase from which star form \citep{Krumholz2019}, here and in our previous paper \citepalias{Crocker2020a},
we seek to cut a broad swathe across the parameter space of star-forming galaxies, and determine where within this parameter space CRs might be important agents of feedback.
We break this task down into two parts.
\citetalias{Crocker2020a} addresses the question:
What fraction of
the total ISM pressure is typically supplied by CRs
as a function of galaxy parameters? 
In other words: How important to the overall gas dynamics in typical star-forming galaxies can CRs be?
In this paper
we use the mathematical set-up of our previous papers 
to
address a rather specific, follow-up question:  
What is the critical flux of cosmic rays above which
a hydrostatic equilibrium within a given column of gas is precluded? 
In other words: At what point do cosmic rays -- accelerated as a result of the star formation process itself -- start to drive outflows in galaxies? We emphasise that we are not addressing the question of whether CRs can re-accelerate or re-heat winds that have been launched by other mechanisms, a question addressed by a number of previous authors as discussed above. Instead, we seek to determine under what conditions it becomes inevitable that the CRs themselves begin to lift neutral interstellar gas out of galactic discs, certainly rendering the neutral gas atmosphere unstable, and
potentially giving rise to a cool galactic wind.

The remainder of this paper is structured as follows: in \autoref{sec:setup} we briefly recap the mathematical setup of the problem and, in particular, write down the ordinary differential equation (ODE) system that describes a self-gravitating gaseous disc that maintains a quasi-hydrostatic equilibrium while subject to a flux of CRs injected at its midplane; in \autoref{sec:equilibria} we present, describe, and evaluate the numerical solutions of our ODEs; in \autoref{sec:implications} we consider the astrophysical implications of our findings for CR feedback on the dense, star-forming gas phase of spiral galaxies; we further discuss our results and summarise in \autoref{sec:discussion}. 

\section{Setup}
\label{sec:setup}

\subsection{Physical Model: Recapitulation}

We provide a detailed description of
the physical system we model 
in the companion paper \citepalias{Crocker2020a}.
In brief, our model is similar
to one previously invoked by us in studies of radiation pressure feedback \citep{Krumholz2012,Krumholz2013, Crocker2018, Crocker2018b, Wibking2018}: an idealised 1D representation of a portion of a galactic disc with total gas mass per unit area $\Sigma_{\rm gas}$ and gas fraction $f_{\rm gas}$, supported by a combination of turbulent motions with velocity dispersion $\sigma$, magnetic fields, and CR pressure, and confined by gravity. CRs (or radiation) are injected into this medium at the midplane with flux $F_{c,0}$. 
In the radiation context we have previously shown that, when the injected radiation flux exceeds a critical value, the system is destabilised and equilibrium becomes impossible. Numerical simulations confirm that radiation-driven winds are possible only in those systems for which equilibria do not exist.
Here we are interested to determine whether a similar critical flux exists for CRs, since, if it does, that would suggest the circumstances under which it is possible for CRs to launch outflows of material out of galactic discs.

\subsubsection{Equations for transport and momentum balance}

In \citetalias{Crocker2020a} we provide a detailed derivation of a pair of coupled ordinary differential equations (ODEs)
that describe hydrostatic equilibrium and transport of CRs with losses.
We present only a sketch of this development here for convenience, and refer readers to \citetalias{Crocker2020a} for the full derivation. We treat CRs in the relativistic, fluid dynamical limit whereby they behave as a  fluid of adiabatic index $\gamma_c = 4/3$.
Our ODEs express how the CR pressure and the gas column change as a function of our single
variable, $z$, the height above the midplane.
CRs are assumed to be injected by supernova explosions occurring solely in a thin layer near $z=0$;  in the context of establishing a stability limit, this assumption turns out to be conservative (even though it is not realistic for most galaxies). We show in \citetalias{Crocker2020a} that the system can be described in terms of four dimensionless functions, $s(\xi)$, $r(\xi) = ds/d\xi$, $p_c(\xi)$, and $\mathcal{F}_c(\xi)$, which represent the dimensionless gas column, gas density, CR pressure, and CR flux as a function of dimensionless height $\xi$. These functions are prescribed by two equations. The first is the dimensionless CR transport equation,
\begin{equation}
   \label{eq:PcrSqrdDimless}
   \frac{\tau_{\rm stream}}{\beta_s} \frac{d \mathcal{F}_c}{d\xi} = 
   -  \tau_{\rm abs}  r p_c  +  \tau_{\rm stream} \frac{dp_c}{d\xi},
\end{equation}
where $\tau_{\rm stream}$, $\tau_{\rm abs}$, and 
$\beta_s$ are all defined below and
\begin{equation}
 \label{eq:Fc}
    \mathcal{F}_c = - \frac{\beta_s}{\tau_{\rm stream}} r^{-q} \frac{dp_c}{d\xi}
\end{equation}
is the dimensionless CR flux expressed  
in the standard diffusion approximation \citep{Ginzburg1964}\footnote{Note that we can describe the process in terms of diffusion even if the microphysical transport process is predominantly streaming, as long as we are averaging over scales comparable to or larger than the coherence length of the magnetic field -- see \citet{Krumholz2019} for further discussion.}, in which $q$ specifies the running of the diffusion coefficient with density (i.e., the diffusion coefficient is proportional to $\rho^{-q}$). The term on the LHS of \autoref{eq:PcrSqrdDimless} represents the gradient of the CR flux, while the two terms on the RHS represent, respectively,
collisional and streaming losses of the CRs\footnote{Note that here we assume that second-order Fermi reacceleration is negligibly small or actually zero; cf.~\citet{Zweibel2017}.}.
The coupled ODE expressing hydrostatic balance is
\begin{equation}
\frac{dp_c}{d\xi} + \phi_{\rm B} \frac{dr}{d\xi} = -\left(1-f_{\rm gas}\right)r - f_{\rm gas} s r.
\label{eq:HydroDimless}
\end{equation}
The terms in \autoref{eq:HydroDimless} are, from left to right, the pressure gradient due to CRs, the pressure gradient due to combined turbulence (treated as isotropic) plus magnetic support\footnote{Note that we thus implicitly assume negligible thermal pressure.}, the gravitational acceleration due to stellar gravity, and the acceleration due to gas self-gravity. 

The dimensionless variables are related to the physical quantities as follows. The dimensionless height is the physical height measured in units of the turbulent scale height:
\begin{eqnarray}
\xi & \equiv & \frac{z}{z_*}
\end{eqnarray}
where 
\begin{equation}
    z_* \equiv \frac{\sigma^2}{g_*},
\end{equation}
(in which the turbulent velocity dispersion of the gas $\sigma$ is assumed constant) and 
\begin{equation}
    g_* = 2 \pi G \frac{\Sigma_{\rm gas}}{\fg}.
\end{equation}
Similarly $s(\xi)$ is the
(dimensionless) fraction of the total half column
contributed by gas in the height range from 0 to $\xi z_*$, and $p_c(\xi)$ is the dimensionless CR pressure obtained by normalizing the dimensional CR pressure to the characteristic midplane pressure $P_*$  (with related energy density $u_* = (3/2) P_*$) given by
\begin{equation}
P_* = g_* \rho_* z_* = \rho_* \sigma^2 = \frac{\pi G}{\fg}  \Sigma_{\rm gas}^2 
\simeq 0.57 \frac{\Sigma_{\rm gas,1}^2}{\fg} \ {\rm eV \ cm}^{-3},
\label{eq:defnPstar}
\end{equation}
where we have defined  
$\Sigma_{\rm gas,1}=\Sigma_{\rm gas}/(10\,M_\odot\,\mathrm{pc}^{-2})$
and
\begin{equation}
\rho_* \equiv \frac{\Sigma_{\rm gas}}{2 z_*}
\end{equation}
is the characteristic matter density. The local density as a function of height is $\rho_* (ds/d\xi)$.

Other parameters appearing in the coupled ODEs are as follows:
The coefficients
$\tau_{\rm abs}$ and $\tau_{\rm stream}$ appearing on the RHS of 
\autoref{eq:PcrSqrdDimless}
are, respectively, the optical depths of the gas column to CR absorption and scattering (see \autoref{eq:tausDefn} and \autoref{eq:tauaDefn} below).
In \autoref{eq:HydroDimless}, 
$\phi_B$ on the LHS lies in the range 0 to 2
and specifies the importance of magnetic effects in modifying the pressure due solely to gas turbulence (with values $>1$ indicating magnetic pressure support and values $<1$ indicating confinement by magnetic tension).
We adopt $q=1/4$ and $\phi_B = 73/72$ as fiducial values, but our results are only weakly sensitive to these choices -- for further discussion see \citetalias{Crocker2020a}.

The cosmic ray optical depth parameters are given by
\begin{eqnarray}
\label{eq:tausDefn}
\tau_{\rm stream} & = & \frac{\beta_s}{K_* \beta} \\
\tau_{\rm abs} & = & \frac{1}{K_* \beta} \tau_{\rm pp}.
\label{eq:tauaDefn}
\end{eqnarray}
where $\beta_s \equiv v_s/c$ 
denotes the dimensionless CR streaming speed, $\beta\equiv \sigma/c$ is the dimensionless ISM velocity dispersion, 
and $K_*$ is the dimensionless midplane CR diffusion coefficient  expressed in units of the effective diffusion coefficient for convective transport:
\begin{equation}
\kappa_* = K_* \kappa_{\rm conv},
\label{eq:K_defn}
\end{equation}
where
\begin{eqnarray}
    \kappa_{\rm conv} & = & \frac{z_* \sigma}{3} =   \frac{\sigma^3 \ \fg}{6 \pi \ G \ \Sigma_{\rm gas}} \nonumber \\
    & \simeq & 3.8 \times 10^{26} \ \rm{cm}^2 \ \rm{s}^{-1} \ \sigma_1^3 \ \fg \ \Sigma_{\rm gas, 1}^{-1},
    \label{eq:kappa_conv}
\end{eqnarray}
and we have defined $\sigma_1 = \sigma/10$ km s$^{-1}$. Note that, as convection sets a lower limit 
to the rate of diffusion, $K_* \geq 1$. We apply this limit to all of the CR transport models we describe below.
The $\tau_{\rm pp}$ parameter
appearing in  the definition of $\tau_{\rm abs}$
is the optical depth for ``absorption''
of cosmic rays 
via the hadronic collisions they experience
in the limit of rectilinear propagation at speed $c$ from the midplane
to infinity through (half of the total)
gas column $\Sigma_{\rm gas}$:
\begin{equation}
\tau_{\rm pp} = \frac{\Sigma_{\rm gas}}{2 \Sigma_{\rm pp}} 
\label{eq:tauppDefn}
\end{equation}
where 
\begin{equation}
\Sigma_{\rm pp} \equiv \frac{\mu_p m_p}{3 \eta_{\rm pp} \sigma_{\rm pp}} \simeq \frac{33 {\rm \ g \ cm}^{-2}}{(\eta_{\rm pp}/0.5) (\sigma_{\rm pp}/40 \ {\rm mbarn}) } \simeq 1.6 \times 10^5 \ \msun/{\rm pc}^2\,
\label{eq:Sigmapp}
\end{equation}
is the grammage required to decrease the CR flux by one $e$-folding; here $m_p$ is the proton mass, $\mu_p\simeq 1.17$ is the number of protons per nucleon for gas that is 90\% H, 10\% He by number, $\eta_{\rm pp}$ and $\sigma_{\rm pp}$
are the inelasticity and total cross-section
for hadronic collisions experienced by relativistic CR protons.

The system formed by \autoref{eq:PcrSqrdDimless} and \autoref{eq:HydroDimless} is fourth-order, and thus requires four boundary conditions. Two of these apply to the density, and are 
\begin{eqnarray}
\label{BC_1}
s(0) & = & 0 \\ 
\lim_{\xi\to\infty} s(\xi) & = & 1, \label{BC_2}
\end{eqnarray}
which amount to asserting that the gas half column is zero at the midplane, and that $\lim_{z\to\infty} \Sigma_{\rm gas,1/2}(z) =  1/2 \ \Sigma_{\rm gas}$. The remaining two apply to the CR pressure and flux, and are
\begin{equation}
\frac{\tau_{\rm stream}}{\beta_s}
\mathcal{F}_c(0) =
\frac{1}{K_*\beta}  \frac{F_{c,0}}{F_*} \equiv f_{\rm Edd},
 \label{BC_3}
\end{equation}
at $\xi = 0$ and
\begin{equation}
\label{BC_4}
\lim_{\xi \to \infty}  \mathcal{F}_c = \lim_{\xi \to \infty} 4 \beta_s p_c.
\end{equation}
The first of these, \autoref{BC_3}, is set by the CR flux $F_{c,0}$ entering the gas column; here $F_* = c P_*$ is the scale flux for our non-dimensional system, and $f_{\rm Edd}$ is the Eddington ratio, which gives the ratio of the momentum flux carried by the CRs to that imparted by gravity. The second, \autoref{BC_4}, asserts that the CR flux approach the value for free-streaming as $\xi\to\infty$. Again, we refer readers to \citetalias{Crocker2020a} for a full derivation of these conditions.

\subsection{CR transport models}
\label{ssec:CR_microphysics}

To complete the specification of the system, we require expressions for $K_*$ and $\beta_s$, the normalised CR diffusion coefficient and streaming speed. These depend on the microphysics of CR confinement, and here we consider the same three models for this process as in \citetalias{Crocker2020a}. These are:

\subsubsection{Streaming (fiducial case)}

We are interested in the feedback effects of CRs on the (predominantly) neutral ISM, which at the midplane of a galactic disc constitutes $\sim 50\%$ of the volume \citep{Dekel2019}, and close to 100\% in the densest starbursts \citep{Krumholz2019}, and the vast majority of the mass. Thus our fiducial case is for CR transport through such a medium. As discussed in \citetalias{Crocker2020a} and shown in \citet{Krumholz2019}, in such a medium strong ion-neutral damping prevents interstellar turbulence from cascading down to the small scales of CR gyroradii, which are the only scales that efficiently scatter CRs. Thus the only disturbances in the magnetic field with which CRs interact are those they themselves generate via the streaming instability. Thus CRs stream along field lines, but for the relatively low (but still relativistic) CR energies that dominate the CR energy budget, streaming instability limits the streaming speed to the ion Alfv\'en velocity of the medium,
\begin{equation}
    v_{A,i} = \frac{\sigma}{\sqrt{2\chi} M_A},
\end{equation}
where $M_A$ is the Alfv\'en Mach number of the Alfv\'enic turbulence modes in the ISM and $\chi$ is the ionisation fraction by mass. For a dynamo-generated field $M_A \approx 1-2$ (\citealt{Federrath14a, Federrath16a}, \citetalias{Crocker2020a}), and astrochemical models show that the ionisation fraction $\chi$ ranges from $\sim 10^{-2}$ in Milky Way-like galaxies with relatively diffuse neutral media \citep{Wolfire2003} to $\sim 10^{-4}$ in dense starbursts \citet{Krumholz2019}. On larger scales, the CR diffusion coefficient is therefore set by the combination of streaming at this speed along the field lines, and the random walk of the field lines themselves in the turbulence. For this model, we show in \citetalias{Crocker2020a} that
\begin{eqnarray}
    \label{eq:Kstar_str}
    K_* & = & \frac{1}{\sqrt{2\chi} M_A} \\
    \beta_s & = & \frac{\beta}{\sqrt{2 \chi} M_A} \\
    \label{eq:tau_stream_str}
    \tau_{\rm stream} & = & M_A^3 \\
    \label{eq:tau_abs_str}
    \tau_{\rm abs} & = & \frac{\sqrt{2\chi} M_A^4}{\beta} \tau_{\rm pp},
\end{eqnarray}
and we use $q = 1/4$ as our fiducial choice as introduced above\footnote{This corresponds to the physical limit where the turbulent velocity dispersion is density independent, there is a local turbulent dynamo acting, and the ionization fraction becomes independent of the local gas density; while this latter is unlikely to hold strictly, as we have previously shown (\citetalias{Crocker2020a} and \citealt{Krumholz2019}), our results are not strongly dependent on $q$ so long as $0<q<1$.}.
For a given choice of $M_A$ and $\chi$, and a galactic disc of specified $\Sigma_{\rm gas}$ and $\sigma$ (which set $\tau_{\rm pp}$ and $\beta$, respectively), these expressions complete the specification of the system.

\subsubsection{Scattering}

Our second model is based on the premise that, although we are interested in feedback on the neutral ISM, ionised gas nevertheless fills $\sim 50\%$ of the midplane volume in most galaxies \citep{Cox1974, Dekel2019}, with the fraction rising as one goes away from the midplane, and thus CR transport might take place predominantly in the ionised phase of the ISM. Indeed, {\it in situ} observations suggest that such is the case for the local CR population seen at Earth 
\citep[e.g.,][]{Ghosh1983,Jones2001}. In this case CRs may still interact predominantly with their own self-generated turbulence, in which case we return to a situation much like the streaming model, except with $\chi = 1$. The more interesting possibility, therefore, is that, although CRs do stream at speed $v_s = v_{A,i} = \sigma/\sqrt{2} M_A$, they also scatter off turbulence that is part of the large-scale turbulent cascade in the ISM, and that this scattering is what sets the diffusion coefficient. In this case, we show in \citetalias{Crocker2020a} that transport coefficients are given by
\begin{eqnarray}
    \label{eq:Kstar_scat}
    K_* & = & \frac{1}{\beta} \left(\frac{G}{2\fg}\right)^{p/2} \left(\frac{E_{\rm CR} M_A}{e\sigma^2}\right)^p \\
    \beta_s & = & \frac{\beta}{\sqrt{2} M_A} \\
    \label{eq:tau_stream_scat}
    \tau_{\rm stream} & = & \frac{3 \beta}{\sqrt{2} M_A^2} 
    \left(\frac{E_{\rm CR}}{e \sigma^2} \sqrt{\frac{G}{2 f_{\rm gas}}}\right)^{-p}
    \\
    \label{eq:tau_abs_scat}
    \tau_{\rm abs} & = & \frac{3 \tau_{\rm pp}}{M_A} \left(\frac{E_{\rm CR}}{e \sigma^2} \sqrt{\frac{G}{2 f_{\rm gas}}}\right)^{-p}.
\end{eqnarray}
Here $E_{\rm CR}$ is the CR energy (we adopt $E_{\rm CR}=1$ GeV $\equiv E_{\rm CR,0}$ as a fiducial choice), $e$ is the elementary charge, and $p$ is the index of the turbulent power spectrum -- $p=1/3$ (corresponding to $q = 1/6$) for a Kolmogorov spectrum, and $p=1/2$ (corresponding to $q = 1/4$, i.e., the fiducial value) for a Kraichnan spectrum, though which value of $p$ we choose makes little difference to the qualitative results. As with the streaming model described above, for a particular choice of $E_{\rm CR,0}$, $p$, and $M_A$, the above expressions allow us to compute the transport coefficients $\beta_s$, $\tau_{\rm stream}$, and $\tau_{\rm abs}$ for any choice of galactic disc parameters $\Sigma_{\rm gas}$ and $\sigma$. Compared to the streaming model, the scattering model generally predicts smaller streaming optical depths in all galaxies, and comparable diffusion rates and absorption optical depths in Milky Way-like galaxies. The models differ mainly in their predictions for denser and more rapidly star-forming galaxies, where the scattering model predicts slower transport and greater absorption optical depths than in the Milky Way (due to stronger turbulence), while the streaming model predicts the opposite (due to the lower ionisation fraction in denser galaxies allowing faster CR streaming). We provide a more detailed comparison in \citetalias{Crocker2020a}.

\subsubsection{Constant diffusion coefficient}

The third model we consider is a purely empirical one:
The empirically-determined 
diffusion coefficient for $\sim$GeV CRs in the Milky Way is close to $\kappa_{\rm *,MW} \equiv 10^{28}$ cm$^{2}$ s$^{-1}$ \citep[e.g.,][]{Ptuskin2006} and, in this model,
we simply assume $\kappa$ in all galaxies is given by this value. We thus assume that CRs stream through a fully ionised medium, as in the scattering case, but we take the dimensionless diffusion coefficient to be $K_* = \kappa_{\rm *,MW}/\kappa_{\rm conv}$. The corresponding expressions for the dimensionless numbers entering the equilibrium equations are
\begin{eqnarray}
    \label{eq:Kstar_con}
    K_* & = & \frac{6\pi G \Sigma_{\rm gas} \kappa_{\rm *,MW}}{\fg \sigma^3} \\
    \beta_s & = & \frac{\beta}{\sqrt{2} M_A} \\
    \label{eq:tau_stream_con}
    \tau_{\rm stream} & = & \frac{1}{\sqrt{2} M_A} \left(\frac{\kappa_{\rm conv}}{\kappa_{\rm *,MW}}\right) \\
    \label{eq:tau_abs_con}
    \tau_{\rm abs} & = & \frac{\tau_{\rm pp}}{\beta} \left(\frac{\kappa_{\rm conv}}{\kappa_{\rm *,MW}} \right).
\end{eqnarray}
Since we can write down the convective diffusion coefficient $\kappa_{\rm conv}$ as a function of $\Sigma_{\rm gas}$ and $\sigma$ (c.f.~\autoref{eq:kappa_conv}), this again represents a complete specification of the system.

\section{The Cosmic Ray Eddington Limit}
\label{sec:equilibria}

With this review of our dimensionless ODE system, and having dealt with the microphysics of CR transport, we are now in a position to address the basic question posed in this paper: under what conditions does it become impossible for a galactic disc forced by CRs from below to remain hydrostatic? To answer this question, we first describe a numerical method to identify this limit in the space of the dimensionless variables that characterise our system (\autoref{subsec:numerical}), we use this method to obtain critical stability curves in this space (\autoref{subsec:critical_curves}), and then we translate from the space of dimensionless variables to the space of observable galaxy properties (\autoref{subsec:eddington}).

\subsection{Numerical method}
\label{subsec:numerical}

We must solve \autoref{eq:PcrSqrdDimless} and \autoref{eq:HydroDimless} numerically. Because the boundary conditions for the system, \autoref{BC_1} - \autoref{BC_4}, are specified at different locations, the system forms a boundary value problem, which we solve using a shooting algorithm as follows: we have $s(0)=0$ from \autoref{BC_1}, and we start with an initial guess for the mid-plane density $r(0) = s'(0)$ and pressure $p_c(0)$. These choices together with \autoref{BC_3} allow us to compute the midplane CR pressure gradient $p'_c(0)$, so that we now have a set of four initial values at $s=0$ and can integrate outwards until $s(\xi)$ and $p_c(\xi)$ approach constant values at large $\xi$. In general our guess will not satisfy \autoref{BC_2}, i.e., $s(\xi)$ will go to a value other than unity as $\xi\to\infty$. We therefore iteratively adjust $s'(0)$ while holding $p_c(0)$ fixed, until \autoref{BC_2} is satisfied. In general, however, this choice will not obey \autoref{BC_4}, i.e., the CR flux will not go to the correct value as $\xi\to\infty$. We therefore now iteratively adjust our guess for $p_c(0)$. We continue to iterate between our guesses for $s'(0)$ and $p_c(0)$ until the system converges and all boundary conditions are satisfied, or until  convergence fails (see below).

\begin{figure}
\includegraphics[width=\columnwidth]{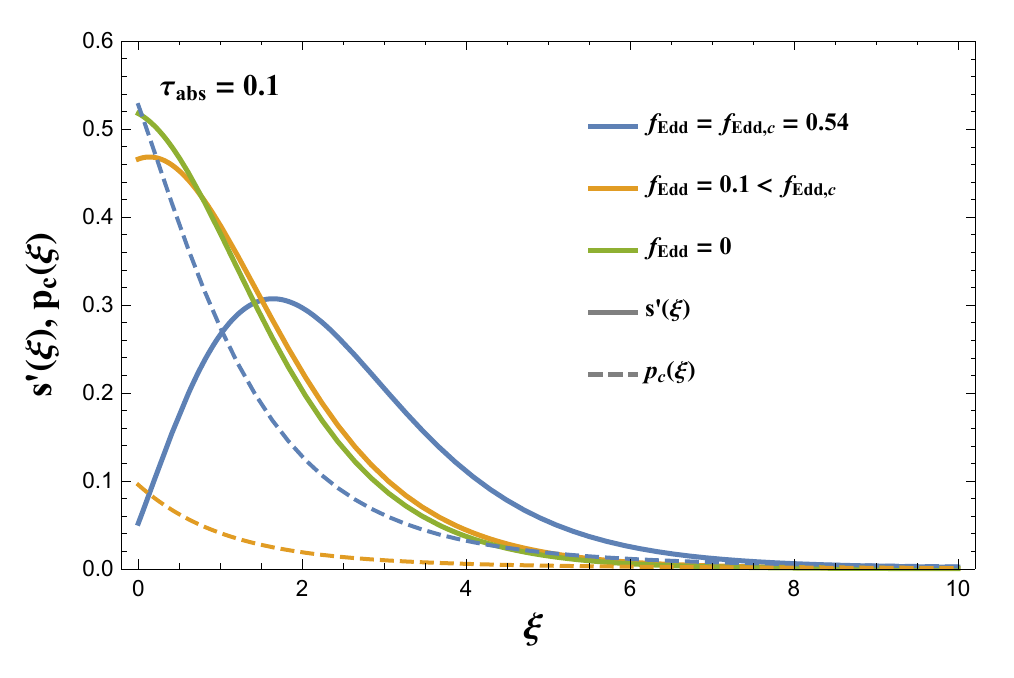}
\includegraphics[width=\columnwidth]{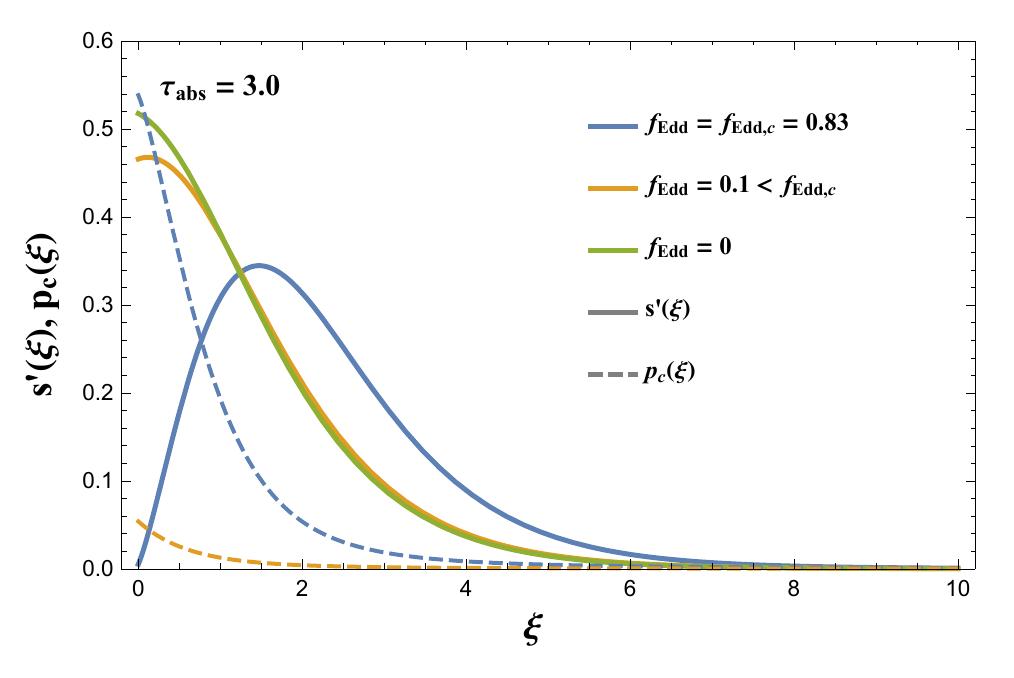}
\caption{Profiles of (dimensionless) volumetric density $r(\xi)=s'(\xi)$ (solid) and (dimensionless) CR pressure $p_c(\xi)$ (dashed) for various representative cases as indicated in the legend. 
The upper panel is for $\tau_{\rm abs} = 0.1$, the lower is for $\tau_{\rm abs} = 3.0$; 
otherwise, parameters common between the panels are 
$q = 1/4, \tau_{\rm stream} = 1, \fg = 0.9, \phi_{\rm B} = 73/72$.
In both panels, 
the blue curves are evaluated for the critical $f_{\rm Edd}$ case and 
the yellow curves are for a sub-critical $f_{\rm Edd}$ value;  the solid green curve is the density profile of a gas column supported purely by turbulence (with $\phi_{\rm B} = 73/72$). Note that because $\sigma$ is constant, $s'(\xi)$ is equivalent to the dimensionless turbulent pressure, and thus the ratio of solid and dashed curves of the same colour is also the ratio of turbulent to CR pressure.
\label{plotDensity}
}
\end{figure}

A crucial feature of solutions to this system is that, as $f_{\rm Edd}$ increases at fixed $\tau_{\rm abs}$ and $\tau_{\rm stream}$, the dimensionless midplane density $s'(0)$ decreases monotonically, approaching zero at a finite value of $f_{\rm Edd}$. We illustrate this behaviour for two example cases in \autoref{plotDensity}. We refer to the value of $f_{\rm Edd}$ for which this occurs as the critical Eddingtion ratio, $f_{\rm Edd,c}$. No solutions exists for $f_{\rm Edd} > f_{\rm Edd,c}$, and thus $f_{\rm Edd,c}$ represents the largest Eddington ratio for which it is possible for a gas column through which CRs are forced to remain in equilibrium. Larger values of $f_{\rm Edd}$ necessarily render the system unstable. Mathematically, this manifests in that we are unable to find values of $s'(0)$ and $p_c'(0)$ such that, when we integrate \autoref{eq:PcrSqrdDimless} and \autoref{eq:HydroDimless}, the resulting solution satisfies the boundary conditions \autoref{BC_2} and \autoref{BC_4} as $\xi\to\infty$. The shooting method fails to converge.

We determine the value of the critical Eddington ratio $f_{\rm Edd,c}(\tau_{\rm abs},\tau_{\rm stream})$ as a function of $\tau_{\rm abs}$ and $\tau_{\rm stream}$ as follows. We start with a small value of $f_{\rm Edd}$, for which a solution is guaranteed to exist because in the limit $f_{\rm Edd}\to 0$,  \autoref{eq:PcrSqrdDimless} and \autoref{eq:HydroDimless} are completely decoupled; the former just reduces to the equation for an isothermal atmosphere, and the latter to a nonlinear diffusion equation with losses, the analytic solution for which is given by \citet{Krumholz2019}. 
We use the shooting procedure described above to obtain the numerical solution for this small value of $f_{\rm Edd}$. We then progressively increase $f_{\rm Edd}$ and solve again, using the solution for the previous value as a starting guess. Eventually we reach a value of $f_{\rm Edd}$ for which the shooting method fails to converge, and no solution exists. Once we find this value, we iteratively decrease and increase $f_{\rm Edd}$ in order to narrow down the value $f_{\rm Edd,c}$ for which a solution ceases to exist. We iterate in this manner until we have determined $f_{\rm Edd,c}$ for a given $\tau_{\rm abs}$ and $\tau_{\rm stream}$ to some desired tolerance. \autoref{plotDensity} confirms that the value of $f_{\rm Edd,c}$ we obtain by this procedure is indeed such that $s'(0)$ is close to zero, although in practice how close we are able to push $s'(0)$ to zero depends on the tolerances we use in our iterative solver -- in the vicinity of $f_{\rm Edd,c}$, the value of $s'(0)$ becomes exquisitely sensitive to $f_{\rm Edd}$. This is visible in the upper panel of \autoref{plotDensity}, where our solution for $f_{\rm Edd} \approx f_{\rm Edd,c}$ has $s'(0)\approx 0.05$, but if we increase $f_{\rm Edd}$ by even 1\%, then solutions cease to exist entirely.

\subsection{Critical curves}
\label{subsec:critical_curves}

\begin{figure}
\includegraphics[width=\columnwidth]{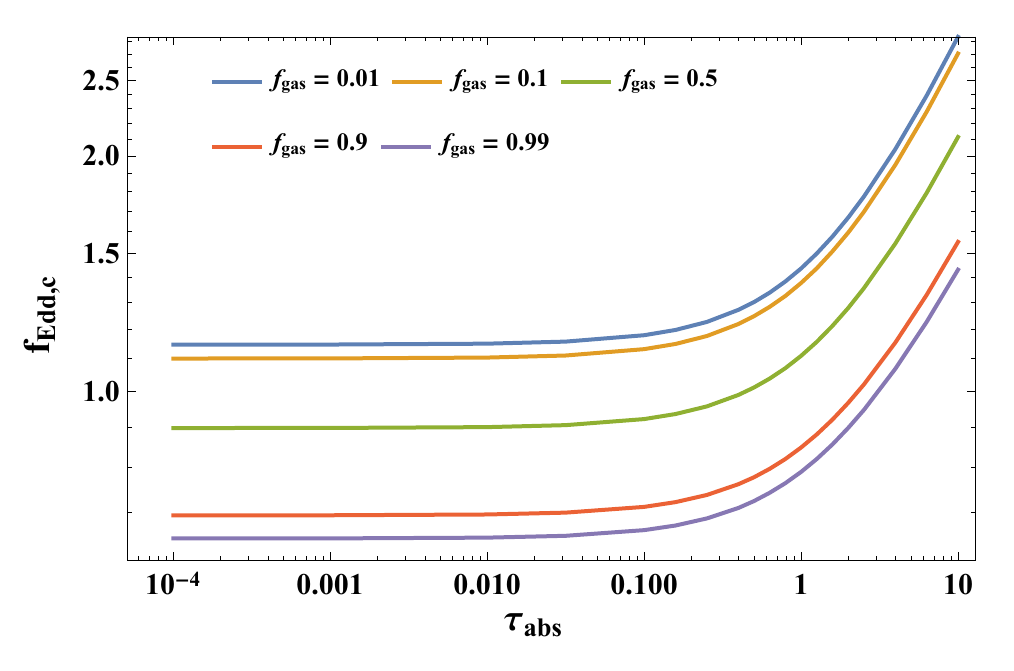}
\includegraphics[width=\columnwidth]{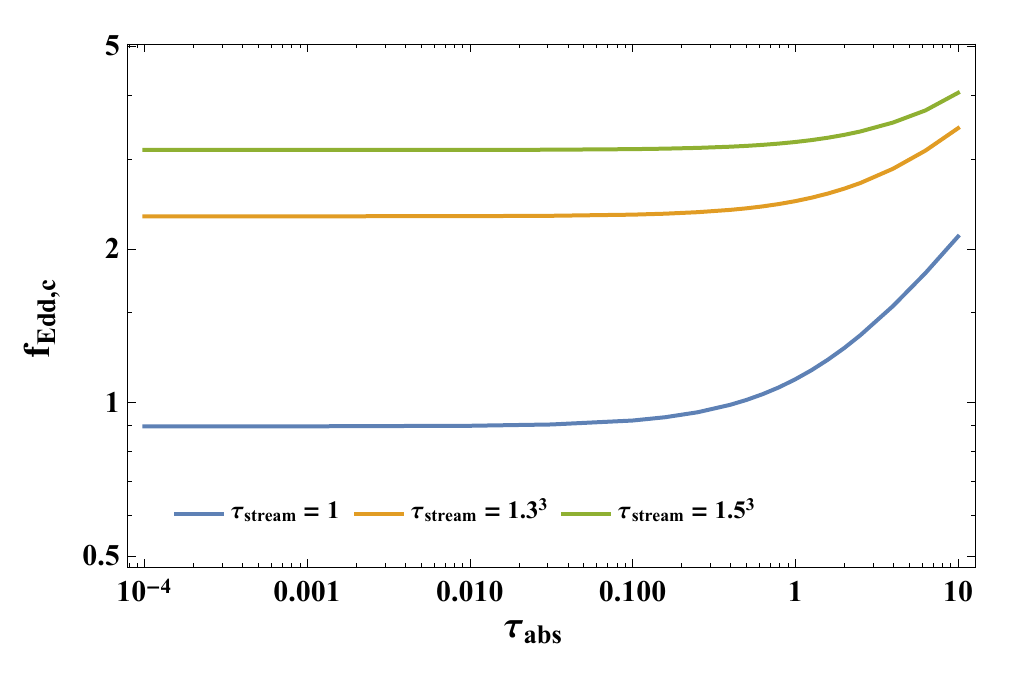}
\caption{{CR Eddington limit $f_{\rm Edd,c}$ as a function of effective optical depth $\tau_{\rm abs}$}.
{\bf Upper panel:} $f_{\rm Edd,c}$ versus $\tau_{\rm abs}$ at fixed $\tau_{\rm stream} = 1$ and for a range of gas fractions $\fg$ as indicated in the legend; a higher $\fg$ renders the column (somewhat) less stable (i.e., reduces $f_{\rm Edd,c}$), while a higher $\tau_{\rm abs}$ renders the column {\it more} stable once we are in the optically thick regime (because hadronic collisions reduce the steady state CR pressure). {\bf Lower panel:} $f_{\rm Edd,c}$ as a function of $\tau_{\rm abs}$ for fixed $\fg = 0.5$ and a range of $\tau_{\rm stream}$ values as indicated in the legend; a higher $\tau_{\rm stream}$ renders the column more stable (again because losses -- in this instance due to the streaming instability -- reduce the steady state CR pressure).
\label{plotVADERPaperCosmicRaysPrelim1}
}
\end{figure}

We show sample values of $f_{\rm Edd,c}$ determined via the procedure described in \autoref{subsec:numerical} in \autoref{plotVADERPaperCosmicRaysPrelim1}; the top panel shows $f_{\rm Edd,c}$ as a function of $\tau_{\rm abs}$ for fixed $\tau_{\rm stream}$ at several values of $f_{\rm gas}$, while the bottom panel shows $f_{\rm Edd,c}(\tau_{\rm abs})$ for fixed $f_{\rm gas}$ at several values of $\tau_{\rm stream}$. Qualitatively, the behaviour of the solution with respect to $\tau_{\rm abs}$ is that, at small $\tau_{\rm abs}$, $f_{\rm Edd,c}$ approaches a fixed, $\cal{O}$(1) value. At large $\tau_{\rm abs}$, we find that $f_{\rm Edd,c}$ begins to scale
$\appropto \tau_{\rm abs}$. We also find that, at low $\tau_{\rm abs}$, we have a rough scaling $f_{\rm Edd,c} \propto \tau_{\rm stream}$ scaling (cf.~lower panel of \autoref{plotVADERPaperCosmicRaysPrelim1}). Finally, we find that increasing $\fg$ renders the column less stable for other parameters held fixed; this is as expected given that gas self-gravity must vanish in the midplane.

We can understand the observed scalings of $f_{\rm Edd,c}$ with $\fg$, $\tau_{\rm abs}$, and $\tau_{\rm stream}$ via some straightforward analytic considerations. Of these, $\fg$ is the simplest. We note that, in the limit $f_{\rm Edd} \ll 1$, the equation of hydrostatic balance (\autoref{eq:HydroDimless}) has the usual solutions $s'(\xi) = e^{-\xi/\phi_B}$ for $\fg = 0$ and $s'(\xi) = \sech^2 (\xi/2\phi_B)$ for $\fg = 1$; at $\xi = 0$, these solutions have $s' = 1/\phi_B$ and $1/2\phi_B$, respectively, so the density at the midplane is twice as high with $\fg=0$ as with $\fg=1$. In between these limits, the midplane density scales as approximately $1/[\phi_B (1+\fg)]$. Since the critical $f_{\rm Edd}$ corresponds to the point where $s'(0)\to 0$, we expect that configurations starting with a larger value of $s'(0)$ at low $f_{\rm Edd}$ should have higher $f_{\rm Edd,c}$, as $f_{\rm Edd,c} \propto \left.s'(0)\right|_{f_{\rm Edd} \ll 1}$. This suggests a scaling $f_{\rm Edd,c} \appropto 1/[\phi_B (1+\fg)]$, which is consistent with our numerical results.

The scalings for $\tau_{\rm abs}$ and $\tau_{\rm stream}$ require only slightly more consideration. We expect our system to approach the critical limit when the midplane CR pressure, $P_c$, becomes significant in comparison to the pressure required to keep the column in hydrostatic equilibrium, $P_*$. In steady state, the midplane CR pressure (or energy density, which differs just by a factor of $3$), in turn, will be set by the product of the CR energy injection rate -- set by star formation -- and a dwell time $t_c$ for CRs injected into the galaxy. Thus we have
\begin{equation}
    \left(\frac{P_c}{P_*}\right)_{z=0} \sim \frac{(F_{c,0}/z_*) t_c}{P_*} = \frac{\sigma t_c}{z_*} K_* f_{\rm Edd},
    \label{eq:PcPstar_oom}
\end{equation}
where $F_{c,0}$ is the energy injected per unit area, and we write the energy injected per unit volume as $F_{c,0}/z_*$ under the assumption that the CRs are distributed over a height of order $z_*$. In the second step, we made use of \autoref{BC_3} to rewrite $F_{c,0}$ in terms of the Eddington ratio.

The dwell time for a CR will be set by the minimum of the time required for it to be lost to a collision, $t_{\rm col}$, to have its energy sapped by streaming losses, $t_{\rm stream}$, or to escape from the galaxy via diffusion, $t_{\rm esc,diff}$:
\begin{equation}
    t_c \sim \left(t_{\rm col}^{-1} + t_{\rm stream}^{-1} + t_{\rm esc,diff}^{-1}\right)^{-1}.
    \label{eq:t_c}
\end{equation}
We can rewrite each of the three ratios appearing inside the parentheses in the above equation in terms of our dimensionless parameters. The collisional loss time is (c.f.~equation 11 of \citetalias{Crocker2020a})
\begin{equation}
    t_{\rm col} \sim \frac{1}{c (\rho_*/\mu_p m_p) \sigma_{\rm pp} \eta_{\rm pp}} \sim \frac{1}{K_* \tau_{\rm abs}} \left(\frac{z_*}{\sigma}\right),
\end{equation}
where we have dropped factors of order unity, and in the second step we have made use of \autoref{eq:tauaDefn} and \autoref{eq:tauppDefn}. Similarly, the streaming loss time is (c.f.~equation 49 of \citetalias{Crocker2020a})
\begin{equation}
    t_{\rm stream} \sim \frac{z_*}{v_s} \sim \frac{1}{K_* \tau_{\rm stream} }\left(\frac{z_*}{\sigma}\right),
\end{equation}
where in the second step we have used \autoref{eq:tausDefn}. Finally, the diffusive escape time is (c.f.~equation 47 of \citetalias{Crocker2020a})
\begin{equation}
    t_{\rm esc,diff} \sim \frac{z_*^2}{\kappa_*} \sim \frac{1}{K_*} \left(\frac{z_*}{\sigma}\right),
\end{equation}
where we have used \autoref{eq:K_defn}. Inserting these factors into \autoref{eq:t_c} for $t_c$, and thence into \autoref{eq:PcPstar_oom}, we find
\begin{equation}
    \left(\frac{P_c}{P_*}\right)_{z=0} \sim f_{\rm Edd} \left(1 + \tau_{\rm abs} + \tau_{\rm stream}\right)^{-1}.
\end{equation}
Thus if we expect the midplane ratio $P_c/P_*$ to be of order unity when $f_{\rm Edd}$ is at the critical value, it follows immediately that
\begin{equation}
f_{\rm Edd,c} \appropto \frac{1 + \tau_{\rm abs} +  \tau_{\rm stream}}{\phi_B(1+\fg)},
\label{eq:fEddRough}
\end{equation}
where we have now re-inserted the scaling with $\phi_B$ and $\fg$ derived above. This is not, of course, an exact expression, but its scalings are qualitatively correct, as we have seen, and account for the following phenomena:
CRs exert a pressure which is i) enhanced by the diffusive nature of their propagation
\citep[c.f.][]{Socrates2008}
-- this leads to the constant term on the RHS --
but attenuated by ii) their collisional  and iii) their streaming  losses; these lead to the $\propto \tau_{\rm abs}$ and $\propto \tau_{\rm stream}$ terms, respectively.

\begin{figure}
\includegraphics[width = \columnwidth]{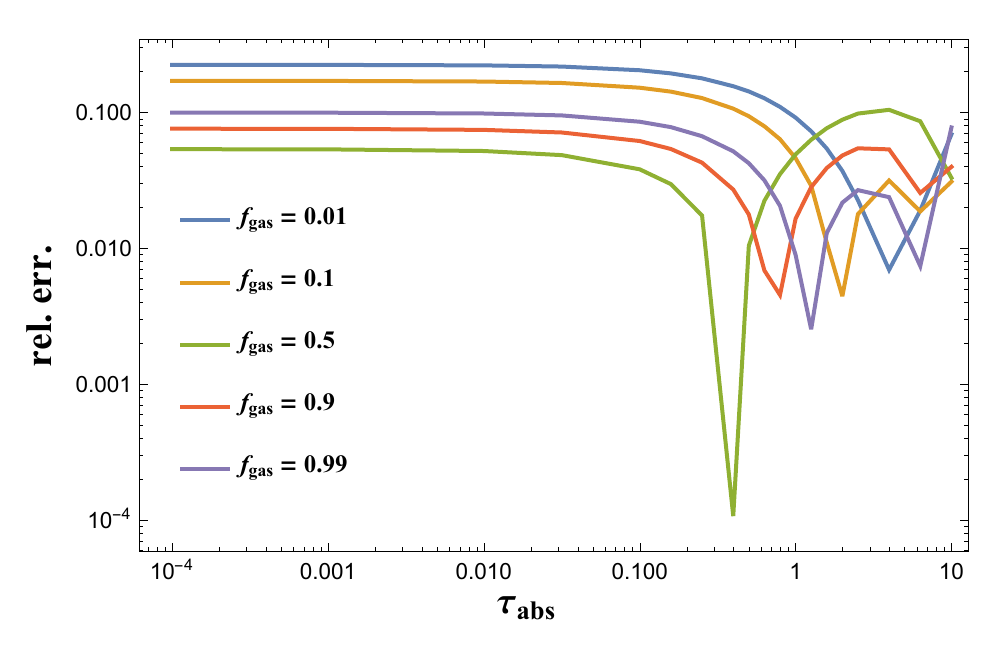}
\includegraphics[width = \columnwidth]{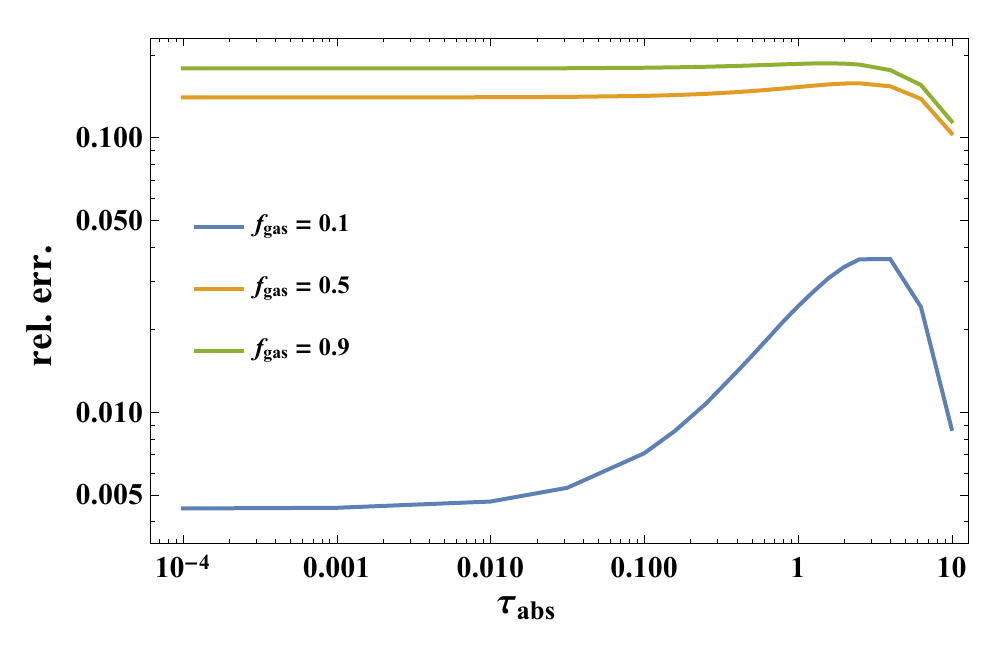}
\includegraphics[width = \columnwidth]{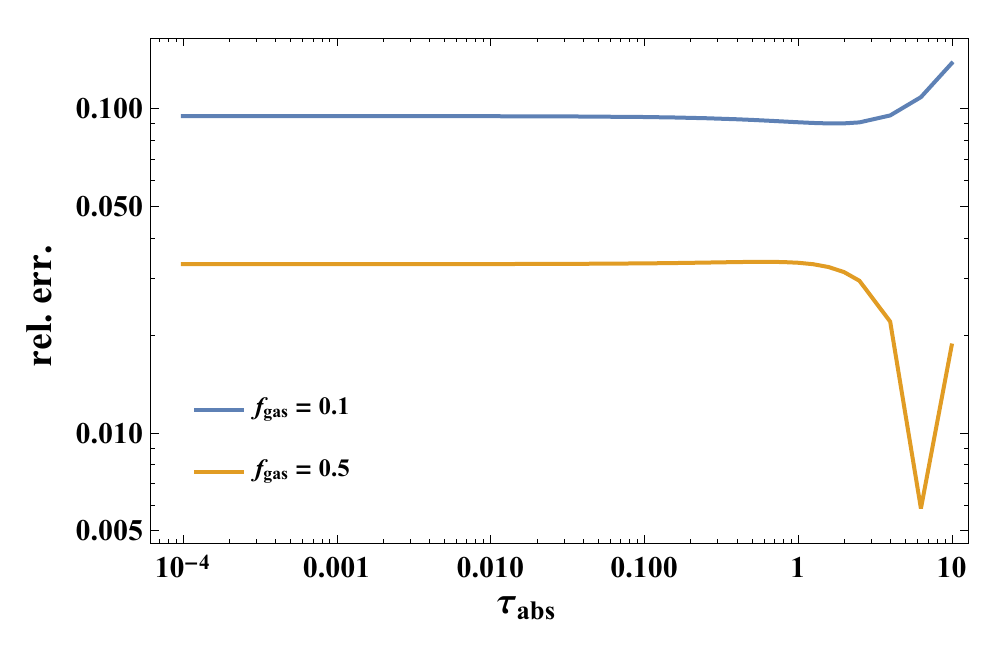}
\caption{
\label{plotRelativeError}
Relative error between the analytic approximation and the full numerically-determined CR critical Eddington ratio. From top to bottom, the panels are for $\tau_{\rm stream} = \left\{1^3,1.3^3,1.5^3 \right\}$; gas fractions are as labelled in each panel's legend.
}
\end{figure}

It will be convenient for the remainder of this paper to use our understanding of the scaling behaviour of $f_{\rm Edd,c}$ to derive an approximate analytic fit that we can use in lieu of the full, numerically-determined solution. We adopt the functional form given by \autoref{eq:fEddRough}, and after some numerical experimentation to find coefficients that minimise the error, we arrive at the approximate relationship
\begin{equation}
    \label{eq:fEddcFit}
    f_{\rm Edd,c} \approx f_{\rm Edd,c,fit} \equiv \frac{1}{2.4 \phi_B (1+\fg)}
     \left(0.26 + \frac{\tau_{\rm abs}}{2.5} + \frac{\tau_{\rm stream}}{0.31}\right) . 
\end{equation}
We plot the relative error in this fit, defined as 
\begin{equation}
    {\rm rel. \ err.} \equiv \frac{\left|f_{\rm Edd,c}- f_{\rm Edd,c,fit}\right|}{f_{\rm Edd,c}},
\end{equation}
in \autoref{plotRelativeError}. The figure demonstrates that our approximation is accurate to $\lsim 20$ percent for $\tau_{\rm abs} < 10$ and $f_{\rm gas} = 0.01 - 0.99$.

\subsection{The cosmic ray stability limit for star-forming galaxies}
\label{subsec:eddington}

The final step in our calculation is to translate our stability limit from dimensionless ($f_{\rm Edd}$, $\tau_{\rm abs}$, $\tau_{\rm stream}$) to the physical variables describing a star-forming galactic disc. In particular, we are interested in the highest star formation rate (since star formation produces supernovae that are the primary source of CRs) that a disc can sustain before it becomes unstable to the development of CR-driven outflows. 

The conversion from dimensionless to physical variables is straightforward. Given a gas surface density $\Sigma_{\rm gas}$, velocity dispersion $\sigma$, gas fraction $\fg$, Alfv\'en Mach number $M_A$, and either (depending on our choice of CR transport model) an ionisation fraction $\chi$ or CR energy $E_{\rm CR}$, we can compute the corresponding $\tau_{\rm stream}$ and $\tau_{\rm abs}$ values from \autoref{eq:tau_stream_str} and \autoref{eq:tau_abs_str} (for our fiducial streaming model), \autoref{eq:tau_stream_scat} and \autoref{eq:tau_abs_scat} (for the scattering transport model), or \autoref{eq:tau_stream_con} and \autoref{eq:tau_abs_con} (for the constant model). From these values plus $f_{\rm gas}$ we can compute the critical Eddington ratio $f_{\rm Edd,c}$ using either the numerical procedure outlined in \autoref{subsec:numerical}, or, with much less computational expense, our approximate fitting formula (\autoref{eq:fEddcFit}).

We can obtain a corresponding star formation rate per unit area $\dot{\Sigma}_\star$ from this as follows. First, following \citetalias{Crocker2020a}, we write the CR flux as
\begin{equation}
\label{eq:Fc0}
    F_{c,0} = \epsilon_{c,1/2} \dot{\Sigma}_\star,
\end{equation}
where $\epsilon_{c,1/2}$ is the energy injected into CRs in each galactic hemisphere per unit mass of stars formed. We adopt a fiducial value $\epsilon_{c,1/2} \approx 5.6\times 10^{47}$ erg $M_\odot^{-1}$, which corresponds to assuming \citet{Chabrier2005} initial mass function, that all stars with mass $\geq 8$ $M_\odot$ end their lives as supernovae with total energy $10^{51}$ erg, and that 10\% of this SN energy is eventually injected into CRs. Substituting \autoref{eq:Fc0} into the definition of $f_{\rm Edd}$ (\autoref{BC_3}), we have
\begin{eqnarray}
    \label{eq:Sigma_star_fEdd}
    \dot{\Sigma}_\star & = & \frac{F_*}{\epsilon_{c,1/2}} K_* \beta f_{\rm Edd} = \frac{\pi G}{\epsilon_{c,1/2}} \frac{\Sigma_{\rm gas}^2 \sigma}{\fg} K_* f_{\rm Edd} 
    \nonumber
    \\
    & = & 4.9\times 10^{-4} \, \frac{\Sigma_{\rm gas,1}^2 \sigma_1}{\fg} K_* f_{\rm Edd} \; M_\odot\mbox{ pc}^{-2}\mbox{ Myr}^{-1}.
\end{eqnarray}
By plugging our value of $f_{\rm Edd,c}$ into \autoref{eq:Sigma_star_fEdd}, together with the appropriate value of $K_*$ for our chosen CR transport model (\autoref{eq:Kstar_str}, \autoref{eq:Kstar_scat}, or \autoref{eq:Kstar_con}), we obtain the  critical star formation rate $\dot{\Sigma}_{\star,\rm c}$ above which galaxies become unstable to CRs.

In order to actually plot $\dot{\Sigma}_{\star,\rm c}$ versus $\Sigma_{\rm gas}$, we require values of $\sigma$, $f_{\rm gas}$, and $\chi$, which vary systematically with $\Sigma_{\rm gas}$ on average (e.g., higher surface density galaxies tend to have higher velocity dispersion), but which are not single-valued functions of $\Sigma_{\rm gas}$ either. To avoid a proliferation of curves, we adopt the same strategy as in \citetalias{Crocker2020a}: we interpolate between plausible values of these parameters as a function of $\Sigma_{\rm gas}$. Specifically, we adopt 
\begin{eqnarray}
\label{eq:fg_fit}
\fg\left(\Sigma_{\rm gas}\right) & \equiv & 0.11 \ \Sigma_{\rm gas,1}^{0.32}  \\
\chi\left(\Sigma_{\rm gas}\right) & \equiv & 
0.013 \
\Sigma_{\rm gas,1}^{-0.79}  \\
\sigma\left(\Sigma_{\rm gas}\right) & \equiv & 
8.5  \ 
\Sigma_{\rm gas,1}^{0.39} \ {\rm km/s} \label{eq:sigmav}
\, .
\label{eq:sigma_fit}
    \label{eq:chiPhen}
\end{eqnarray}
We emphasise that these are not intended to be accurate fits; they are simply intended to provide smooth functions we can use to reduce the multidimensional parameter space of $\Sigma_{\rm gas}$, $\sigma$, $\fg$, and $\chi$ to a single dimension so that we can represent it on a plot. 

\begin{figure*}
\includegraphics[width=0.7\textwidth]{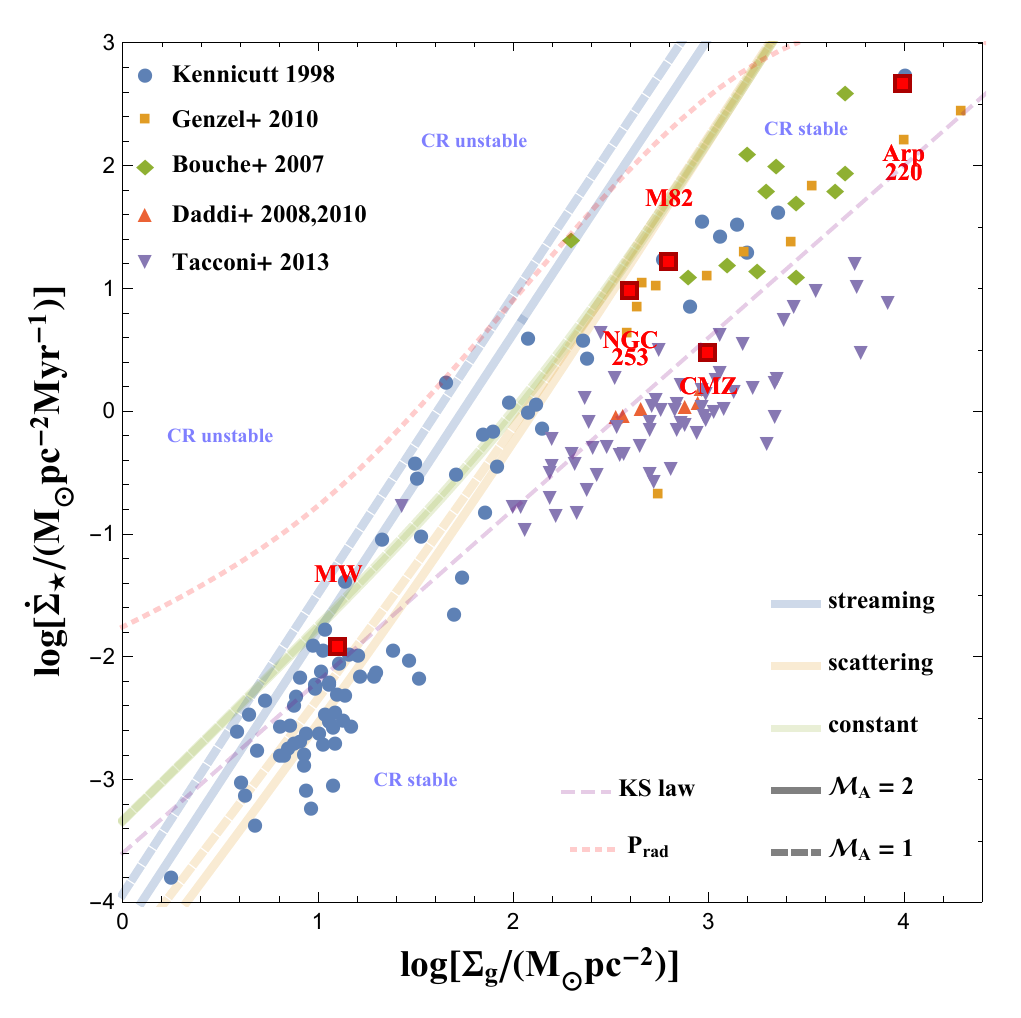}
\caption{
\label{plotVADERPaperCosmicRaysPrelim2}
Thick coloured lines show critical stability curves $\dot{\Sigma}_{\rm \star,c}$, i.e., the star formation rate per unit area at which CR pressure precludes hydrostatic equilibrium, 
computed assuming one of
three different CR transport modes
as indicated (with streaming constituting our fiducial model), and for $\mathcal{M}_A = 2$ (solid)
and $\mathcal{M}_A = 1$ (dashed)
together with fiducial parameter choices for all models,
and using the fits for $\fg$, $\chi$, and $\sigma$ as a function of $\Sigma_{\rm gas}$ given by \autoref{eq:fg_fit} - \autoref{eq:sigma_fit}. The dashed, diagonal, purple line is the \citet{Kennicutt1998} star formation scaling. 
The dotted red line is the 
critical star formation rate surface density obtained for radiation pressure feedback.
This curve smoothly interpolates from the
single-scattering case (adopting a
critical Eddington ratio 0.6 on the basis of the calculation by \citet{Wibking2018}  and
assuming a fixed flux-mean dust opacity per mass of dust–gas mixture of $\kappa = 1000$ cm$^2$/g) into the regime where the atmosphere is optically thick to FIR radiation due to the dust opacity \citep[][]{Crocker2018} with a cross-over at $\Sigma_{\rm gas} \simeq 10^3 \ \msun/$pc$^2$ (and we have assumed a young, $< 7$ Myr old, stellar population for the optically thick part of the curve).
Finally, points show observations drawn from the following sources: local galaxies from \citet{Kennicutt1998}, z $\sim 2$ sub-mm galaxies from \citet{Bouche2007}, and galaxies on and somewhat above the star-forming main sequence at z $\sim 1-3$ from \citet{Daddi2008,Daddi2010b,Genzel2010,Tacconi2013}. The red data points show the Solar neighborhood (`MW') and the Central Molecular Zone (`CMZ') of the Milky Way, and three, local starbursts whose $\gamma$-ray emission is modelled in \citet{Krumholz2019}. The observations have been homogenised to a \citet{Chabrier2005} IMF and the convention for $\alpha_{\rm CO}$ suggested by \citet{Daddi2010a}; see \citet{Krumholz2012b} for details. 
}
\end{figure*}

With this understood, we plot $\dot{\Sigma}_{\star,\rm c}$ as a function of $\Sigma_{\rm gas}$ in \autoref{plotVADERPaperCosmicRaysPrelim2}. We show curves for the cases of i) our fiducial streaming model for CR transport (blue), ii) the alternative scattering (yellow), and iii) the case of constant $\kappa$ (green); for all of these modes we show both results for both $\mathcal{M}_A = 2$ (solid) and $\mathcal{M}_A = 1$ (dashed).
For any particular curve, the  stable region is below and to the right, while the unstable region is above and to the left.
\autoref{plotVADERPaperCosmicRaysPrelim2} also shows a selection of observed galaxies culled from the literature (see \citetalias{Crocker2020a} for details of the data compilation), with some particularly significant galaxies
shown by the red points: the Milky Way datum (`MW'), its Central Molecular Zone (CMZ), and the nearby starbursts NGC253, M82, and Arp 220, whose $\gamma$-ray emission we modelled in \citet{Krumholz2019}.

\section{Implications}
\label{sec:implications}

\autoref{plotVADERPaperCosmicRaysPrelim2} is the central result of this paper. Here we discuss its implications, and explore the physical origin of the result and its sensitivity to a variety of assumptions and parameter choices that we have made.

\subsection{For which galaxies can CRs drive outflows from the star-forming ISM?}

We start by examining our fiducial CR trasnport model, indicated by the blue lines in \autoref{plotVADERPaperCosmicRaysPrelim2}. An immediate conclusion we can draw is that, for physically-plausible scalings of the parameters, the CR stability curve  patrols a region very close to the top of the occupied part of the $(\Sigma_{\rm gas},\dot{\Sigma}_\star)$ plane for star-forming galaxies with low gas surface densities typical of the Galaxy and local dwarfs. This correspondence strongly suggests that CR feedback 
on the neutral gas may be an important mechanism in such galaxies: it might limit the ability of galaxies to make excursions above the locus where most of them like to be, or it might be responsible for launching winds and ejecting gas in galaxies that do wander upwards to higher star formation rates.

Conversely, it is evident that, at the higher gas surface densities
encountered in local starbursts and high-redshift star-forming galaxies, {\it all} the critical curves (not just the one for our fiducial model) diverge away from the observed distribution of galaxies. This implies
that CRs cannot drive winds in these systems \citepalias[cf.][]{Crocker2020a}. Mathematically, such a divergence must occur for the following reason: from \autoref{eq:Sigma_star_fEdd}, for the range of $\tau_{\rm abs}$ for which $f_{\rm Edd,c} \sim$ const at fixed $\tau_{\rm stream}$ (cf. \autoref{plotVADERPaperCosmicRaysPrelim1}) we have shown that the critical star formation rate surface density scales approximately as
$ \dot{\Sigma}_{\rm \star,c} \appropto \tau_{\rm abs}^2 \propto \Sigma_{\rm gas}^2$. On the other hand, the observed surface density of star formation rises with gas surface density with an index $<2$. Physically, the divergence occurs because the high gas number densities in starburst systems kill CRs quickly,  meaning that the energy density/pressure they represent cannot build up to be anything comparable to hydrostatic pressures \citep[cf][]{Lacki2011,Thompson2013,Crocker2020a}. On the other hand, such a situation constitutes a recipe for CR calorimetry, so these systems are expected -- indeed, directly inferred, in a limited number of cases -- to be good hadronic $\gamma$-ray sources \citep[cf.][]{Torres2004,Thompson2007,Lacki2010,Lacki2011,Yoast-Hull2016,Peretti2019,Krumholz2019}.

\subsection{Instability of galaxies under scattering}

A second significant point that is evident from \autoref{plotVADERPaperCosmicRaysPrelim2} is that, for the cases of both {\it constant} diffusivity and {\it scattering}, the critical curves cut well into the occupied
region of parameter space for lower
surface gas density galaxies; many such galaxies, including the Milky Way, are unstable under this scenario. Since the constant model is not physically well-motivated, this is not particularly surprising; the more surprising result is for the scattering model. We remind readers that
this model for CR transport applies in an environment where there is an extrinsic turbulence cascade that reaches down to the gyroradius scale of the energetically-dominant $\sim$GeV CRs. We have shown \citet{Krumholz2019} that this is not the case for the neutral medium that dominates the mass and forms the stars. On the other hand, the long-standing classical interpretation of the totality of the CR
and diffuse gamma-ray emission phenomenology is that the spectrum of the Milky Way's steady state, hadronic cosmic ray distribution is informed by exactly this process of scattering on extrinsic turbulence \citep[e.g.,][]{Jones2001}.

This is not necessarily a contradiction: unlike the situation for starbursts \citep{Krumholz2019}, the MW midplane ISM is not single phase.
Rather, the filling factors of the dense, neutral phase and the more diffuse, ionised phase are similar in the midplane \citep[][]{Kalberla2009}. Moreover, the ionised gas filling factor increases towards unity as we rise away from the midplane and individual SNRs or stellar cluster superbubbles form chimneys into the hot, ionised halo.
Thus some -- potentially large -- fraction of CRs accelerated by SNR shocks in Milky Way-like conditions may never encounter a large grammage of matter in escaping the midplane, i.e., they experience an effective $\tau_{\rm abs} \ll 1$ and, incidentally, also $\tau_{\rm stream} \ll 1$ given that streaming losses are generically small, in relative terms, for the scattering mode. These CRs -- provided that the classical  picture of scattering on an extrinsic turbulent cascade is roughly correct -- will render the ionised gas column hydrostatically unstable.
This applies for Milky Way conditions according to \autoref{plotVADERPaperCosmicRaysPrelim2}.
Of course, an assumption here is that the classical picture of scattering on extrinsic turbulence is essentially the correct one for the ionised phase, and this may not actually hold
\citep[e.g.,][]{Zweibel2017,Blasi2019}.
On the other hand, recognising that the ionised gas column for most low surface-density galaxies will only constitute some $\lsim$10\% of the total column, \autoref{plotVADERPaperCosmicRaysPrelim2} may actually tend to exaggerate stability with respect to cosmic ray feedback in the ionised phase for such galaxies.

In summary, as has long been recognised \citep{Jokipii1976,Ko1991,Breitschwerdt1991,Everett2008,Socrates2008}, it is hard to escape the conclusions that for local galaxy conditions, CRs will likely drive winds in the ionised gas phase.
This will lead to mass loss over cosmological timescales.
However, none of these considerations preclude the existence of a hydrostatic equilibrium in the dense, neutral phase that allows it to sustain the star formation process.
Overall, the picture we thus arrive at here is that
there are effectively two  transport regimes
operating for CRs in Milky Way-like galaxies
(according to the ISM phase within which CRs are propagating).
Qualitatively this agrees with 
the long-standing argument \citep{Ginzburg1980}
that the correct 
interpretation of local CR phenomenology\footnote{Specific aspects 
of CR phenomenology that support the existence of a CR halo include i) the very low levels of CR anisotropy and ii)
the difficulty encountered in otherwise reconciling
CR age measurements obtained with unstable ``clock" nuclei (like
$^{10}$Be) with the grammage encountered by typical $\sim$GeV+ CRs as inferred from secondary to primary CR nuclei ratios.} 
is that
there are distinct disc and halo CR propagation zones, with the halo diffusion coefficient significantly (i.e., $3-10 \ \times$) larger  than the disc one (while the characteristic matter density in the disc  is, of course, substantially larger than that in the halo).

\subsection{The role of the Alfv\'enic Mach number}

One of the important parameters that appears in our models is the Alfv\'enic Mach number of the turbulence. This parameter controls the streaming speed and thus the strength of streaming losses directly, and also affects the overall diffusion rate -- weakly for the scattering or constant models, strongly for the streaming transport model. We have argued based on dynamo theory that $M_A$ will always be in the range $\sim 1-2$ in galactic discs \citep{Federrath14a, Federrath16a}, but it is important to investigate to what extent our conclusions are dependent on this argument. 

\begin{figure}
\includegraphics[width=\columnwidth]{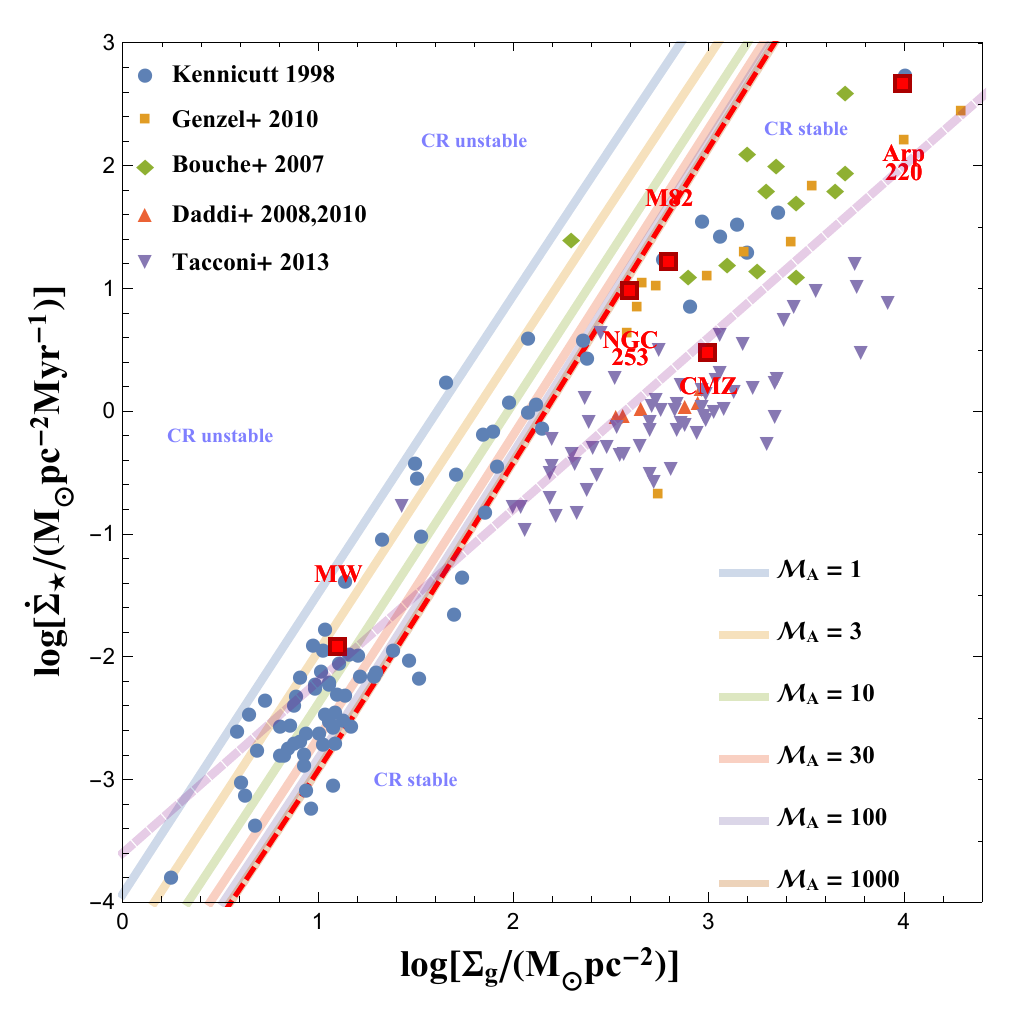}
\caption{
\label{plotVADERPaperCosmicRaysKSplaneMultiMA}
Critical stability curves 
for the streaming mode of CR transport with $M_A$ as given in the legend.
Parameters and data points are identical to those used in \autoref{plotVADERPaperCosmicRaysPrelim2}, except for $M_A$.
The red, dashed curve shows the stability
curve in the high $M_A$ limit, where it is completely determined by hadronic losses. 
}
\end{figure}

We first investigate this in the context of the streaming model, where the $M_A$-dependence is greatest. In \autoref{plotVADERPaperCosmicRaysKSplaneMultiMA}
we show critical stability curves for this model computed with various value of $M_A$. In the high $M_A$ limit the stability curve becomes universal
(i.e., independent of the transport mode)  because the CR energy density is set purely by hadronic losses right across the range of gas surface density; the critical curve in this limiting case is shown as the dashed red line in the figure.
Note that, were ordinary, local disc galaxies operating in this limit, their neutral ISM phase would be unstable and driving a strong outflow, something that we do not observe.
This implies, minimally, that magnetic fields in such galaxies are not too far below equipartition with respect to turbulent energy density (as expected in the case that a local turbulent dynamo is operating, and as we see directly in the Milky Way).
More speculatively, it may be that gas motions induced by CRs in the case that the atmosphere is CR unstable  help to drive magnetic fields towards equipartition at the lower surface gas density end of the distribution. 

The difference between $M_A=1$ and 2 is far smaller for the scattering and constant models. This occurs because, for these modes, the streaming loss timescales are relatively long while the diffusive escape or collisional timescales are either mildly dependent or completely independent of $M_A$.
The critical curves for the scattering and constant $\kappa$ cases also become identical at high $\Sigma_{\rm gas}$.
This occurs because the  collisional timescale formula is universal and collisional losses are solely responsible for setting the CR energy density in this regime. The streaming curve does not exhibit this behaviour because here
streaming losses remain comparable to collisional losses even up to very high $\Sigma_{\rm gas}$. The streaming case approaches the streaming and constant curves at high $\Sigma_{\rm gas}$ only in the limit $M_A\gg 1$.

\subsection{Unimportance of convective transport}

On the basis of \autoref{eq:Sigma_star_fEdd},
$\dot{\Sigma}_{\rm \star,c} \propto K_* f_{\rm Edd,c}$, so
it might seem that we should expect that
there should be clear structures in the critical curves where $K_*$ becomes equal to unity and the CR transport mechanism changes between convection and some other process. The $K_* = 1$ regime occurs for $\Sigma_{\rm gas} \lesssim 10$ $M_\odot$ pc$^{-2}$ for the streaming transport model, and for $\Sigma_{\rm gas} \gtrsim 10^3$ $M_\odot$ pc$^{-2}$ for the starburst and scattering transport models, yet there are clearly no sharp features visible in the critical curves. Indeed, even if we explicitly ignore convection (i.e., we allow $K_* < 1$ when computing $\dot{\Sigma}_{\rm \star,c}$), we find critical curves that are essentially identical.

We can understand the reason for this both mathematically and physically. Mathematically, for a scattering or constant $\kappa$ model of CR transport, the diffusive transport rate only becomes smaller than the convective rate in galaxies with very high surface densities. However, these galaxies also have $\tau_{\rm abs} \gg 1$, and in this limit we have $f_{\rm Edd,c} \propto \tau_{\rm abs}$. Since $\tau_{\rm abs} = \tau_{\rm pp} / K_* \beta$, we arrive at $\dot{\Sigma}_\star \propto \tau_{\rm pp}/\beta$, with no dependence on $K_*$; this is why a sharp change in the value of $K_*$ does not generate a corresponding sharp change in $\dot{\Sigma}_\star$. Physically, the origin of this behaviour is that the only galaxies in which the rate of diffusive CR transport becomes smaller than the rate of convective transport are those with very high gas densities, and thus high $\tau_{\rm abs}$. In these galaxies, the dominant CR loss process is no longer diffusive escape, it is pion production. In this regime, the rate of transport -- diffusive or convective -- is irrelevant to setting the CR energy density. Instead, the CR energy density is simply set by the competition between injection and pion loss, and it is the balance between these two processes that determines the location of the critical curve, the point at which the CR pressure becomes too high to permit hydrostatic balance.

For the streaming model, the transition to convective transport happens at low gas surface density. However, in these galaxies we also have $\tau_{\rm stream} \gtrsim 1$, and thus $f_{\rm Edd,c} \propto \tau_{\rm stream}$. Since $\tau_{\rm stream} \propto 1/K_*$, the dependence of $\dot{\Sigma}_\star$ on $K_*$ again disappears, and thus there is again no sudden change in the critical curve when we reach the convective limit. The physical origin of the behaviour in this place is that, for this transport model, CRs are lost primarily to streaming rather than to escape. We again therefore have a situation where the CR energy density is set by the balance between streaming and injection, a balance that does not depend on the effective diffusion coefficient.

\subsection{Cosmic rays versus radiation pressure as wind launching mechanisms}

CRs are of interest as a feedback mechanism partly because they are much less efficiently lost to cooling than hot gas produced by SN explosions. However, a second appealing aspect of CR feedback is that CRs are ``cool'', in that they can accelerate gas without the need for a shock, and thus naturally explain the presence of low-temperature species in galactic winds. It is therefore interesting to compare CRs to radiation pressure, which is another cool feedback mechanism. Radiation pressure can be delivered either by the direct stellar radiation field or by radiation that has been absorbed by dust and reprocessed into the infrared (cf. red dotted line in \autoref{plotVADERPaperCosmicRaysPrelim2} which interpolates between these limits). At galactic scales, the latter mechanism is only important in the densest starbursts \citep{Thompson2005, Crocker2018, Crocker2018b}, precisely where we have shown that CR feedback is ineffective. The more interesting comparison is therefore in the regime of low surface density galaxies, where direct, or ``single scattering'', radiation pressure dominates \citep[e.g.,][]{Scoville2001, Murray2005,Fall2010,Andrews2011,Thompson2015,Skinner2015, Thompson2016,Wibking2018}. 

For a region of a galactic disc with areal star formation rate $\dot{\Sigma}_\star$, the momentum per unit area per unit time delivered by the radiation field per hemisphere in the single scattering limit is
\begin{equation}
    \dot{\Pi}_{\rm ss} = \frac{1}{c} \dot{\Sigma}_\star \Phi_{1/2},
\end{equation}
where 
\begin{equation}
\Phi_{1/2} \simeq 6.0 \times 10^{50} \ {\rm erg}/\msun
\end{equation}
is the efficiency for conversion of gas mass into radiation (into one galactic hemisphere) via the star formation process \citep{Kennicutt2012}. Note that we are assuming that all of the direct stellar radiation is absorbed. This is an upper limit, but cannot be wrong by a large factor, since, as pointed out by \citet{Andrews2011}, $\sim 1/3$ of the radiation momentum budget is in ionising photons, which will be absorbed even by a tiny column of neutral gas. By comparison, the momentum per unit area per unit time in the upward direction delivered by CRs into each hemisphere, integrating over the gas column, is
\begin{equation}
    \dot{\Pi}_c = -P_* \int_0^\infty \frac{dp_c}{d\xi} d\xi.
\end{equation}
To evaluate the integral, we can make use of \autoref{eq:PcrSqrdDimless}:
\begin{eqnarray}
    \int_0^\infty \frac{dp_c}{d\xi} d\xi & = & \frac{1}{\tau_{\rm stream}} \left[ \int_0^\infty \left(\tau_{\rm abs} r p_c + 
    \frac{\tau_{\rm stream}}{\beta_s}
    \frac{d\mathcal{F}_c}{d\xi}\right) d\xi\right] \nonumber \\
    & = & \frac{1}{\beta_s}\left[\mathcal{F}_c(\infty) - \mathcal{F}_c(0) \right]
    + \frac{f_{\rm Edd} \ f_{\rm cal}}{\tau_{\rm stream}},
\end{eqnarray}
where
\begin{equation}
    f_{\rm cal} = \frac{\tau_{\rm abs}}{f_{\rm Edd}} \int_0^\infty r p_c \, d\xi
\end{equation}
is the ``calorimetric fraction'', i.e., the fraction of all CRs that are lost to pion production (c.f.~equation 64 of \citetalias{Crocker2020a}), and $\mathcal{F}_c(\infty)$ is the flux at $\xi=\infty$.

Therefore we find that the ratio of CR to single-scattering radiation momentum is
\begin{eqnarray}
    \frac{\dot{\Pi}_c}{\dot{\Pi}_{\rm ss}} & = & \left(\frac{\epsilon_{c,1/2}}{\Phi_{1/2}}\right) \frac{1}{\beta_s} \left[1 - f_{\rm cal} - \frac{\mathcal{F}_c(\infty)}
    {\mathcal{F}_c(0)}\right] \nonumber \\
    & \simeq & 9.3\times 10^{-4} \frac{1}{\beta_s} \left[1 - f_{\rm cal} - \frac{\mathcal{F}_c(\infty)}{{\mathcal{F}_c(0)}}\right] \nonumber \\
    & \simeq & \frac{2.8}{v_{s,2}}  
    \left[1 - f_{\rm cal} - \frac{\mathcal{F}_c(\infty)}{{\mathcal{F}_c(0)}}\right]
    \label{eq:cr_vs_ss}
\end{eqnarray}
where $v_{s,2}\equiv v_s/(100$ km/s) and we have made use of \autoref{eq:tausDefn},  \autoref{BC_3}, \autoref{eq:Fc0} to simplify. This expression has a straightforward physical interpretation. The leading numerical factor of $\approx 10^{-3}$ 
in the second line
represents the ratio of energy injected into photons versus energy injected into CRs. The second term, $1/\beta_s$, which is always $\gg 1$, accounts for the fact that CRs transfer momentum to the gas much more efficiently than photons, due to the fact that their propagation speed is limited to a value $\ll c$ by scattering off Alfv\'en waves. Finally, the factor in square brackets just represents the reduction in CR momentum transfer due to loss of CRs by pion production (the $f_{\rm cal}$ term) and due to the escape of some fraction of the injected CRs from the disc without interaction (the $\mathcal{F}_c(\infty)/\mathcal{F}_c(0)$ term). It approaches unity if all of the CR energy is lost to streaming, and becomes smaller if there is significant CR energy loss into other channels. This term can be evaluated numerically from our solutions, and, for the low surface density galaxies with which we are concerned here, is generally in the range $\sim 0.1 - 1$ -- see Section 4.3 of \citetalias{Crocker2020a}.

The implication of \autoref{eq:cr_vs_ss} is that CRs deliver more momentum to the gas than single-scattering radiation pressure if the CR streaming speed satisfies $v_s \lesssim  100$ km s$^{-1}$; the exact condition will depend on details such as the fraction of photon momentum that is actually absorbed, which is likely close to unity in spiral galaxies, but below unity in dust-poor dwarfs. Regardless of the exact numerical limit on $v_s$, the condition is certainly met if the CRs propagate through ionised gas, for which the streaming speed is nearly equal to the total gas Alfv\'en speed, which, for $M_A\sim 1$, is comparable to the $\sim 10$ km s$^{-1}$ velocity dispersion in the ISM. Thus in the scattering or constant $\kappa_*$ CR propagation scenarios, CRs are more important than photons. 

For our favoured streaming scenario, the question of whether CRs or photons are more important is more subtle because the streaming speed in this case is close to the ion Alfv\'en velocity, which, in a weakly-ionised medium, is much larger than the total Alfv\'en velocity or the velocity dispersion. Since we are concerned here with low surface density galaxies whose interstellar media are predominantly atomic, we expect the ionisation fraction $\chi \sim 10^{-2}$ \citep{Wolfire2003}, and thus the ion Alfv\'en speed to be $\approx 10$ times the bulk gas velocity dispersion. This suggests that CRs and single scattering radiation are of roughly comparable importance, and both may contribute to the launching of galactic winds in such galaxies (cf. \autoref{plotVADERPaperCosmicRaysPrelim2}). CRs are probably somewhat more important than photons in low-metallicity dwarfs, where the absence of dust will render galaxies more transparent and thus reduce $\dot{\Pi}_{\rm ss}$, though only by a factor of $\sim 3$ as noted above; on the other hand, it is possible that the equilibrium ionisation fraction is also slightly higher in low-metallicity dwarfs.\footnote{
Also note that, though the (red, dashed) $P_{\rm rad,ss}$ line in 
\autoref{plotVADERPaperCosmicRaysPrelim2} falls above the locus of points at low surface densities, \citet{Thompson2016} argue that radiation pressure in a turbulent medium will be most important along low-column density sightlines not representative of the mean gas surface density.}
For denser galaxies whose interstellar media are largely molecular, $\chi$ is smaller, and the ion Alfv\'en speed correspondingly larger. In these galaxies photons deliver more momentum than CRs; however, this changeover likely has little practical importance, since both direct photons and CRs are generally unimportant in these galaxies.

Finally, we note that our \autoref{eq:cr_vs_ss} is somewhat different from the analogous expression (their equation 21) of \citet{Socrates2008}. We discuss the reasons for this difference in \aref{app:cr_vs_ss}.

\section{conclusions}
\label{sec:discussion}

In this paper we analyse the stability of the neutral, star-forming phase of galactic discs against cosmic ray (CR) pressure.
We
use an idealised model where such discs are 
taken to be plane-parallel slabs of gas confined by stellar and gas self-gravity, and supported by a combination of turbulent and CR pressure. Such a system is characterised primarily by three dimensionless numbers: the effective optical depths of the disc to CR absorption (via $\pi$ production) and to CR streaming, and the CR Eddington ratio (defined by the ratio of the CR momentum flux to the gravitational momentum flux). The primary result of our analysis is that such a system possesses a stability limit: for a given effective optical depth, there exists a maximum CR Eddington ratio above which the system cannot remain hydrostatic. While the nature of the non-linear development of the resulting instability is uncertain, studies of the analogous instability driven by radiation suggests that the result is likely to be a outflow that removes mass until the system is driven back below the stability limit.

Given standard estimates for the efficiency with which SNe inject CRs into galaxies, together with characteristic numbers describing the magnetohydrodynamic turbulence in the ISM and a model for CR transport in a turbulent medium, we can translate our stability limit directly into a line in the space of gas surface density and star formation rate, the so-called Kennicutt-Schmidt (KS) plane. 
We find that the stability limit projected on to the KS plane is close to a line of slope 2, which, for our favoured model of CR transport closely matches the upper envelope of observed systems with the surface densities characteristic of modern spiral and dwarf galaxies, $\Sigma_{\rm gas} \lesssim 300$ $M_\odot$ pc$^{-2}$. While a scaling $\propto \Sigma_{\rm gas}^2$ for the critical star formation rate density with gas surface density is generic to feedback
mechanisms \citep[e.g.][]{Andrews2011}, the fact that our calculation should have produced such a coincidence between the {\it normalization} of the critical curve
and the upper range of the occupied KS parameter space
is surprising: In the dimensionless parameter space of $\tau_{\rm stream}$, $\tau_{\rm abs}$, and $f_{\rm Edd}$ that defines our system, the critical value of $f_{\rm Edd}$ above which the gas column is rendered hydrostatically unstable follows purely from the mathematical form of our ODEs. 
The only astrophysical inputs required to map this to the KS plane are then fundamental constants (e.g., the $pp$ cross-section), quantities describing general physical processes that are unrelated to galaxies (e.g., the saturation field strength of turbulent dynamos), and quantities describing microphysical processes such as the conversion efficiency from supernova kinetic energy to CR energy. The only complex modeling needed is that required to estimate the ionization fraction, which is determined at least partly by the CRs themselves. Given these inputs, the overall similarity of the CR stability limit to the observed galaxy distribution seems unlikely to be a coincidence.
We suggest that the star-forming gas in modern and/or low surface gas density galaxies is
poised close to instability such that 
rather small changes in ISM parameters imply the launching of CR-driven outflows;
CRs -- possibly in concert wtih direct radiation pressure -- thus define the upper limit to the star-formation efficiency of ordinary, star-forming disc galaxies.

In contrast, we find that galaxies with higher gas and star formation surface densities lie well below the CR stability limit. This divergence between the CR stability line and the sequence occupied by observed galaxies has two related but distinct causes. The first is simply that the fundamental scaling that governs all considerations of feedback: the self-gravitational pressure of a galactic disc rises as the square of the gas surface density, whereas the available energy input from star formation, given that the observed index of the Kennicutt-Schmidt relation is $<2$, rises more slowly. However, this alone would not be enough to prevent CRs from becoming significant at high surface densities, since, in the absence of loss mechanisms, CRs would also become increasingly well-confined in high surface density galaxies, and this would cause a superlinear rise in the CR pressure. Indeed, it was precisely this consideration that led \citet{Socrates2008} to conclude that CR feedback dominates in high surface-density galaxies. That it does not do so is due to the second factor that suppresses CR feedback in gas-rich galaxies: the increasing importance of hadronic losses. We show that the critical Eddington ratio above which CRs destabilise a galactic disc scales as the sum of the optical depths of a galactic disc to streaming and hadronic losses. While the former varies only weakly across the star-forming sequence, the latter becomes very large in high surface-density galaxies. Consequently, despite the fact that the discs are starburst galaxies that confine CRs quite well, hadronic losses prevent the CR energy density from building up to the point where CRs are able to launch 
outflows.
Thus we conclude that CRs cannot be dynamically important in the star-forming ISM phase of these galaxies. Conversely, however, due to the importance of pion losses, these galaxies are good CR calorimeters and, therefore,
$\gamma$-ray sources
\citep[cf.][]{Torres2004,Thompson2007,Lacki2010,Lacki2011,Yoast-Hull2016}.

In future work we intend to explore the consequences of the picture set out here and in \citet{Krumholz2019} and \citetalias{Crocker2020a} for understanding the far infrared--radio continuum correlation and the emerging far infrared--$\gamma$-ray correlation, and
to delimit the possible contribution  of hadronic $\gamma$-ray emission from star-forming galaxies to the isotropic $\gamma$-ray flux as predicted by our model.

\section*{Data Availability Statement}

No new data were generated or analysed in support of this research.

\section*{Acknowledgements}

This research was funded by the Australian Government through the Australian Research Council, awards FT180100375 (MRK) and DP190101258 (RMC and MRK). RMC gratefully acknowledges conversations with Felix Aharonian, Geoff Bicknell, Yuval Birnboim, Luke Drury, Alex Lazarian, Chris McKee, Christoph Pfrommer, Heinz V{\"o}lk, and Siyao Xu. MRK and TAT acknowledge support from the Simons Foundation through the Simons Symposium Series ``Galactic Superwinds: Beyond Phenomenology'', during which some aspects of this work were planned. TAT thanks Brian Lacki and Eliot Quataert for discussions and collaboration. TAT is supported in part by National Science Foundation Grant \#1516967 and NASA ATP 80NSSC18K0526.








\appendix

\section{Cosmic rays versus radiation: comparison to the results of Socrates et al.~(2008)}
\label{app:cr_vs_ss}

Our result for the ratio of CR to single-scattering radiation momentum imparted to the gas, \autoref{eq:cr_vs_ss}, is substantially different at first glance from that derived by \citet{Socrates2008}. In this appendix we explain the reasons for this difference. Using our notation, the basic result from \citeauthor[][their equation 21]{Socrates2008} is
\begin{equation}
    \dot{\Pi}_c \sim \tau_{\rm CR} \epsilon_{c,1/2} \dot{\Sigma}_\star,
\end{equation}
where $\tau_{\rm CR}$ is the effective optical depth of the galactic disc to CR scattering, which \citeauthor{Socrates2008} argue is $\sim 10^3$. This expression differs from our \autoref{eq:cr_vs_ss} in that dimensionless factor on the right hand side is $\tau_{\rm CR}$, rather than $[1 - f_{\rm cal} - \mathcal{F}_c(\infty)/f_{\rm Edd}]/\beta_s$.

The difference in the two expressions can be explained by noting that the expression of \citeauthor{Socrates2008} does not incorporate any loss mechanisms for CRs, either streaming or hadronic\footnote{Note that \citeauthor{Socrates2008} do consider CR losses elsewhere in their manuscript (see, for instance, their Appendix C).}. Thus they are here implicitly taking the limits $\tau_{\rm abs} \to 0$ and $\tau_{\rm stream} \to 0$. We can first verify that, if we adopt the same limit, our results reduce to theirs. In this case we cannot use \autoref{eq:cr_vs_ss} directly, because in this limit $f_{\rm cal} \to 0$, $\mathcal{F}_c(\infty)\to f_{\rm Edd}$, and $\beta_s\to 0$, and thus the numerator and denominator of the equation both approach zero. However, for the case of zero losses, \autoref{eq:PcrSqrdDimless} immediately implies $d\mathcal{F}_c/d\xi = 0$, so $\mathcal{F}_c = \mathcal{F}_c(0) = f_{\rm Edd}$ is constant. We then have, from \autoref{eq:Fc},
\begin{equation}
    \int_0^\infty \frac{dp_c}{d\xi} d\xi = -f_{\rm Edd} \int_0^\infty r^q \, d\xi = -f_{\rm Edd} \frac{r(0)^{q+1}}{q+1},
\end{equation}
where 
$q$ is the index describing the scaling of the diffusion coefficient with the ambient density and
$r(0)$ is the value of $r$ at $\xi = 0$, and we have taken $r\to 0$ as $\xi\to\infty$. The quantity $r(0)^{q+1}/(q+1)$ is of order unity, and thus we recover, in dimensional terms
\begin{equation}
    \dot{\Pi}_c \sim P_* f_{\rm Edd}.
\end{equation}
If we now rewrite $f_{\rm Edd}$ in terms of the injected CR flux $F_{c,0}$ using \autoref{BC_3} and \autoref{eq:K_defn}, and dropping factors of order unity, we arrive at
\begin{equation}
    \dot{\Pi}_c \sim F_{c,0} \frac{z_*}{\kappa_*}.
\end{equation}
The quantity $\kappa_*/z_*$ has units of velocity, and can be thought of as the effective velocity which which CRs diffuse, which is lower than the true microphysical velocity by a factor of $\tau_{\rm CR}$. Thus if we further assume that CRs have a microphysical speed of $c$ in between scatterings, then it immediately follows that
\begin{equation}
    \dot{\Pi}_c \sim \tau_{\rm CR} \frac{F_{c,0}}{c},
\end{equation}
which is exactly the \citeauthor{Socrates2008} result.

With this understood, we can now explain why \citeauthor{Socrates2008}'s results differ from our \autoref{eq:cr_vs_ss}. In the absence of losses, the CR pressure that can build up inside the disc is limited only by considerations of hydrostatic equilibrium. If one considers only CR transfer, then for a sufficiently small value of the CR diffusion coefficient $\kappa_*$ (or its dimensionless analog $K_*$), the Eddington ratio $f_{\rm Edd}$ can become arbitrarily large, allowing $\dot{\Pi}_c$ to become similarly large. However, it is not self-consistent to retain the assumptions that $\tau_{\rm stream} \sim 0$ and $\tau_{\rm abs} \sim 0$ as $K_* \to 0$ -- from \autoref{eq:tausDefn} and \autoref{eq:tauaDefn}, we see that $\tau_{\rm stream}$ and $\tau_{\rm abs}$ both scale as $1/K_*$. Thus if a galactic disc has small $K_*$, possibly allowing a large CR pressure to build up, it necessarily also has large $\tau_{\rm stream}$ and $\tau_{\rm abs}$, which reduce or counteract that buildup. Mathematically, this effect manifests in the fact that \autoref{eq:cr_vs_ss} has a coefficient of $[1 - f_{\rm cal} - \mathcal{F}_c(\infty)/f_{\rm Edd}]/\beta_s$, which approaches \citeauthor{Socrates2008}'s factor $\tau_{\rm CR}$ as $\tau_{\rm abs}\to 0$ and $\tau_{\rm stream}\to 0$, but is smaller outside of these limits. Physically, the effect is that, if one attempts to confine CRs by making their diffusion slow, then at the same time this raises the importance of streaming and hadronic losses, which set limits on the extent to which the CR pressure can build up.

\bsp	
\label{lastpage} 
\end{document}